\documentstyle[12pt]{article}

\def\hybrid{\topmargin -20pt  \oddsidemargin 0pt
      \headheight 0pt   \headsep 0pt
      \textwidth 6.25in % A4 paper
      \textheight 9.5in % A4 paper
      \marginparwidth .875in
      \parskip 5pt plus 1pt   \jot = 1.5ex}

\hybrid

\begin{document}

\def\x{\times}
\def\ra{\rightarrow}
\def\beq{\begin{equation}}
\def\eeq{\end{equation}}
\def\beqa{\begin{eqnarray}}
\def\eeqa{\end{eqnarray}}

\sloppy
\newcommand{\be}{\begin{equation}}
\newcommand{\eq}{\end{equation}}
\newcommand{\ov}{\overline}
\newcommand{\un}{\underline}
\newcommand{\p}{\partial}
\newcommand{\la}{\langle}
\newcommand{\bl}{\boldmath}
\newcommand{\ds}{\displaystyle}
\newcommand{\nl}{\newline}
\newcommand{\th}{\theta} 
\renewcommand{\thesection}{\arabic{section}}
\renewcommand{\theequation}{\thesection.\arabic{equation}}
\parindent1em
%\textwidth14.5cm
%\textheight23.0cm
%\oddsidemargin0.5cm
%\topmargin-1.4cm
%\addtocounter{section}{1}
%\setcounter{section}{1}
%\addtocounter{section}{1}
%\newcommand{\resetcounter}{\setcounter{equation}{0}}     
% set counter to zero

\begin{titlepage}
\begin{center}
\hfill IASSNS-HEP-97/139\\
\hfill {\tt hep-th/9801057}\\

\vskip .7in  

{\bf Moduli in N=1 heterotic/F-theory duality}

\vskip .3in

Gottfried Curio$^a$ and Ron Donagi$^b$\footnote{email: 
curio@ias.edu, donagi@math.upenn.edu\\
The first author is partially supported by NSF grant DMS 9627351\\
The second author is partially supported by NSF grant DMS 95-03249}
\\
\vskip 1.2cm

{\em (a) School of Natural Sciences, 
Institute for Advanced Study, Princeton,
NJ 08540 \nl
(b) Department of Mathematics, 
University of Pennsylvania, Philadelphia, PA 
19104-6395}

\vskip .1in

\end{center}

\vskip .2in

\begin{quotation}\noindent

The moduli in a 4D N=1 heterotic compactification on an elliptic CY, 
as well as in the dual F-theoretic compactification, break into "base" 
parameters which are even (under the natural involution of the elliptic 
curves), and "fiber" or twisting parameters; the latter include a 
continuous part which is odd, as well as a discrete part. We interpret all 
the heterotic moduli in terms of cohomology groups of the spectral covers, 
and identify them with the corresponding F-theoretic moduli in a certain 
stable degeneration. The argument is based on the comparison of three 
geometric objects: the spectral and cameral covers and the ADE del Pezzo 
fibrations. For the continuous part of the twisting moduli, this amounts 
to an isomorphism between certain abelian varieties: the connected 
component of the heterotic Prym variety (a modified Jacobian) and the 
F-theoretic intermediate Jacobian. The comparison of the discrete part 
generalizes the matching of  heterotic 5brane / F-theoretic 3brane impurities.

\end{quotation}
\end{titlepage}
\vfill
\eject

{\bf Introduction}\\

The classical string model for $N=1$ supersymmetry
in 4D is the compactification of the heterotic string on a Calabi-Yau $Z$
with a vector bundle $V=V_1+V_2$ which breaks part of the $E_8\times E_8$ 
symmetry. In this paper we describe
the moduli in a 
heterotic compactification on an {\em elliptic} CY, as well as 
in the dual F-theoretic compactification. These moduli
include "base" 
parameters which are even 
(under the natural involution of the elliptic curves), and "fiber" or 
twisting parameters; the latter include a continuous part which is odd,
as well as a discrete part.  
We will interpret all the heterotic moduli in terms of 
cohomology groups of spectral covers, and identify them with the 
corresponding F-theoretic moduli in a certain stable degeneration of the
$K3$ fibers. 
The argument will actually be based on the comparison of three 
geometric objects: the 
spectral and cameral covers and the ADE del Pezzo fibrations.
The full twisting moduli are given, on the heterotic 
side, by a Prym variety constructed from the cameral cover. 
This Prym has a discrete group of components, 
each of which is an abelian variety. 
On the F-theory side the continuous part of the moduli is 
identified with the intermediate Jacobian of the del Pezzo fibration. The 
full moduli are given by the Deligne cohomology, an extension of a 
discrete group by the intermediate Jacobian. We show that the 
two abelian varieties making up the continuous parts are isomorphic, 
in any dimension and for any 
group. As to the discrete parts, we are able to identify the F-theory side 
as a subgroup of finite index in the heterotic side. Assuming that a 
couple of group cohomologies vanish, this becomes an isomorphism 
between the discrete parts as well. The comparison of the discrete 
parts generalizes the matching of  heterotic 5brane / F-theoretic 
3brane impurities.

Let us describe first some background for the problem.
Two aspects of recent progress involve 
the consideration of the associated moduli space [\ref{FMW}], 
[\ref{D}], [\ref{D2}], [\ref{BJPS}], [\ref{FMW2}], and the
occurrence of non-perturbative five-branes in the 
vacuum as correction term in the anomaly cancellation 
\beqa
c_2(Z)=c_2(V_1)+c_2(V_2)+n_5f. \nonumber
\eeqa

This progress was made possible by considering the case
that $Z$ is elliptically fibered $\pi:Z\rightarrow B$
(of fibre $f$ which the five-branes wrap) over a two-fold base $B$. The 
motivation for requiring an elliptic fibration
was twofold: first there exists then a dual $F$-theory 
description by considering a four-fold $X^4$ which is correspondingly
fibered by $K3$ over $B$, thus extending adiabatically over the base $B$
the duality in 8D between the heterotic string on $T^2$ and $F$-theory 
over $K3$ [\ref{V}]. Here the $K3$ is assumed to be elliptically fibered
over ${\bf P}^1$; so this is actually a type IIB compactification on this
${\bf P}^1$ with varying dilaton+RR-scalar $\lambda+ie^{-\phi}$ 
reflecting the complex structure modulus of the elliptic ($F$-theory) 
fibre. The 18 deformations of this type of $K3$ correspond then to
the 16 Wilson lines on the heterotic side + the complex structure
parameter of the (heterotic) elliptic $T^2$ +1 
(the Kahler parameter of the $T^2$)
sharing the moduli space
$SO(2,18;{\bf Z})\backslash SO(2,18;{\bf R})/SO(2;{\bf R})\times 
SO(18;{\bf R})$; the size of the
${\bf P}^1$ corresponds with the heterotic dilaton. For consistency 
of such a four-fold compactification, tadpole cancellation requires
 [\ref{SVW}] that a number $n_3$ of space-time
filling three-branes has to be turned on (we don't turn on gauge bundles
inside the seven-brane [\ref{BJPS}])
\beqa
n_3=\frac{\chi(X^4)}{24}\nonumber
\eeqa
which precise number could then be matched with $n_5$, 
in the case of $V$ a
$G=E_8$ bundle leaving no unbroken gauge group. 
To make the identification
one has to consider that the base
$B^3$, the compactification space of the type IIB string theory with
varying dilaton, is in turn fibered by ${\bf P}^1$ over the base $B^2$
common with the heterotic side, concretely $\eta _{1/2}=6c_1\pm t$ with 
$t=c_1({\cal T})$ the cohomology class characterizing the ${\bf P}^1$ 
fibration of $B^3$ over $B^2$, the ${\bf P}^1$ bundle being given as 
projectivization of the vector bundle ${\cal O}\oplus {\cal T}$ 
(the analogue of the well known relation [\ref{MV}] 
in the 6D case, corresponding
to $B$ now a ${\bf P}^1$, between bundles of instanton number (12+n,12-n)
and $F$-theory for a Calabi-Yau three-fold over the Hirzebruch surface
$F_n$).

The other motivation was that by working with elliptically fibered $Z$
one can adiabatically extend the known results about moduli spaces of 
$G$-bundles over an elliptic curve $E=T^2$, of course taking into account
that such a fiberwise description of the isomorphism class of a bundle 
leaves definitely room for {\it twisting along the base $B$}. The latter
possibility actually involves a two-fold complication: there is a 
continuous as well as a discrete part of these data. It is quite easy to
see this  for  $G=SU(n)$:  in this case $V$ can
be constructed via push-forward of the 
Poincare bundle on the spectral cover $C \times _BZ$, possibly twisted
by a line bundle ${\cal N}$ over the spectral surface $C$ (an n-fold
cover of $B$ (via $\pi$) lying in $Z$), whose first Chern class 
(projected to $B$)
is known from the condition $c_1(V)=0$. So ${\cal N}$ itself is known
up to the following two remaining degrees of freedom: 
first a class in $H^{1,1}(C)$ which projects 
to zero in $B$ (the discrete part), and second an element of 
$Jac(C):=Pic_0(C)$ (the continuous part; the moduli odd under the 
elliptic involution). We will see below how to generalize this to other groups. 

The continuous part is expected
[\ref{FMW}] to correspond on the $F$-theory side to the odd moduli,
related there to the intermediate Jacobian $J^3(X^4)$ of dimension 
$h^{2,1}$, so that the following picture emerges. The moduli space 
${\cal M}$ of the bundles is fibered ${\cal M}\rightarrow {\cal Y}$, with
fibre $Jac(C)$. There is a corresponding  picture on the 
$F$-theory side: ignoring the Kahler classes (on both sides), the moduli 
space there is again fibered. The base is the moduli space of those 
complex deformations which fix a certain complex structure of $Z$; 
the fibre is the intermediate Jacobian 
$J^3(X^4)=H^3(X,{\bf R})/H^3(X,{\bf Z})$
In total,  $h^{2,1}(Z)+h^1(Z,adV)+1=h^{3,1}+h^{2,1}$. (Unspecified Hodge 
numbers refer to $X^4$). The fibre
moduli are 'odd', the deformations belonging to the base 'even'.

The discrete part should correspond to the possibility of turning 
on four-flux in an $M$- respectively $F$-theory compactification, as 
will  be described in more detail below. 
This must, of course,  be included in a description of the fibre
data of ${\cal M}\rightarrow {\cal Y}$. (The continuous and discrete
part together describe
for the four-fold what is known as its Deligne cohomology, which in the 
case considered is an extension of $H^{2,2}(X^4,{\bf Z})$ by $J^3(X^4)$;
here Hodge cohomology with integer coefficients will refer to the obvious
intersection). 

We start in section1  by establishing some notation while briefly reviewing  
the match of the number of the base moduli, coming from ${\cal Y}$.
This is done for $SU(n)$ following [\ref{BJPS}],[\ref{FMW}], and then for 
$E_8$ following [\ref{C}]. We then discuss the match of brane impurities, 
without odd moduli/flux, for $E_8$ [\ref{FMW}] and for $SU(n)$ [\ref{AC}].
Later in the paper we will go on and consider the match 
of brane impurities {\it including} the full twisting degrees of freedom for 
$E_8$. We will also give an interpretation (section 1.2) 
of all the bundle moduli $H^1(Z,ad V)$,
even or odd under the involution, in terms of even respectively 
odd cohomology of the spectral surface , 
including an interpretation of the ${\bf Z}_2$ equivariant 
index of [\ref{FMW}] as
giving essentially the holomorphic Euler characteristic of the spectral
surface.

In section 2 a dictionary is established between the geometry of the
spectral surface and the full $F$-theory moduli, including the discrete 
data. This is again done in two steps: first we show how one can consider
the device of
spectral cover (considered in [\ref{FMW}] for $G$ in the A-series) 
respectively the del Pezzo construction (considered in [\ref{FMW}] 
for $G$ in the E-series) 
when suitably generalized 
as alternative and effectively related descriptions of one and the 
same thing for any bundle. 
 
Our main results concerning the identification of the Prym and the 
Deligne cohomology are obtained in section 3. Our point of view is that
while the bundle of del Pezzos exists only for some groups,
and the spectral cover depends on some non-canonical choices,
there exists in complete generality 
one naturally defined object which we use to relate
the various strands: this is the cameral cover. The reason this is the 
right object is because the distinguished Prym, 
a certain extension of a discrete group by an abelian variety which 
is attached to the cameral cover,  is isomorphic to the group of twisting data.
Having recalled this isomorphism [\ref{D}], we then go on and relate
the distinguished Prym to the analogous groups attached to the
spectral cover and the del Pezzo fibration, 
whose connected components are, respectively, 
the Prym-Tyurin variety and the relative intermediate Jacobian. 

In section 4 we go on to the connection with the $F$-theory side [\ref{MV}]. 
We consider the stable degeneration 
$X^4\rightarrow X^4_{deg}=W_1 \cup_Z W_2$
where the $W_i$ are fibered by del Pezzo surfaces over $B$.
The 8D picture involves a $K3$ degenerating 
into the union of two rational elliptic surfaces 
(here still called del Pezzo and denoted $dP_9$). The base of the
fibration is the union of two projective lines intersecting in a point 
$Q$ over 
which a common elliptic curve $E$ is fibered; roughly speaking
the two $E_8$ contributions in the $K3$ are separated; note that
the transcendental lattice of the $K3$ is $E_8\oplus E_8 \oplus H$, 
with $H$ the 2-dimensional hyperbolic plane piece, which leads to the
18-dimensional space $S:=SO(2,18)/SO(2)\times SO(18)$ divided by 
the appropriate
discrete group; one specializes then to two $E_8$ singularities
at positions $z=0,\infty$ in the $P^1$ base, which after the 'separation'
in two surfaces are again resmoothed; imagine to take 
(for the $dP_9$'s to come alive) the two $f_4,g_6$
parts at $z=0,\infty$ of the original Weierstrass data $f_8,g_{12}$ of 
the $K3$. This corresponds on the 
heterotic side to the large area degeneration of a $T^2$ of the same 
complex structure parameter as $E$ [\ref{FMW}][\ref{AM}]. Intuitively
speaking again, imagine that the $H$ and its counterpart in $S$ above 
correspond to the degrees of freedom represented by 
the complex structure modulus $\tau$ and the area (+ $B$-field) 
modulus $\rho$ of $E$; then in the $\rho \ra i\infty$ limit one finds
in the corresponding boundary component of the quotient 
(discrete$\backslash S$) 
the two
spaces ($W \backslash (E_i\otimes \Lambda _c)$), 'glued' together by 
$\tau (E_1)=\tau (E_2)$, describing the moduli of the two $dP_9$'s
($\Lambda _c$ the coroot lattice of $E_8$, $W$ the Weyl group).
The heterotic invariant $n_5=c_2Z-c_2V_1-c_2V_2$
is then mirrored on the F-theory side by
$n_3=-\frac{\chi(Z)}{24}+\frac{\chi(W_1)}{24}+\frac{\chi(W_2)}{24}$.
Note that, as necessary for the relation to the abelian
variety $Jac(C)$, the relevant 
intermediate Jacobian of the $W_i$ in the stable degeneration
is abelian (note that the intermediate Jacobian of the fourfold $X$ 
itself is also abelian, as the $h^{3,0}(X)=0$). Let us remark further
that we will consider large area for the base $B$ to remain within the
realm of classical geometry. Furthermore we will assume that the unbroken
gauge group is ADE.

We close by pointing out that not
only the occurrence of the four-flux modifies the required number of 
three-branes but that also the three-branes refine in some sense the
quantization condition for the four-flux, excluding for example the
simplest choice of fulfilling the integrality congruence [\ref{W4fl}]
by setting the flux equal to $\frac{1}{2}c_2(X^4)$: just as this would
violate supersymmetry [\ref{BB}] in case of non-primitiveness of $c_2$ it 
violates it also, one sees here, due to the negative number of 
three-branes which would be needed.

\section{Bundle moduli and brane impurities}

We review first some of the setup of [\ref{FMW}] concerning 
the spectral cover construction for V an $SU(n)$ vector bundle.
Then we show how the even respectively odd bundle moduli are reflected in the 
even respectively odd cohomology of the spectral surface. Finally
we collect some results on the comparison of bundle moduli 
and F-theory moduli respectively of the brane impurities for $G=SU(n),E_8$.\\ \\
{\bf 1.1 Spectral cover for $G=SU(n)$}

Let V be an $SU(n)$ vector bundle over the elliptically fibered
heterotic Calabi-Yau threefold $\pi:Z\rightarrow B$ with section $\sigma$
of normal bundle ${\cal L}^{-1}=K_B$. We also think of $\sigma$ as
element of $H^2(Z)$, and then $\sigma|_{\sigma}=-c_1|_{\sigma}$
(unspecified Chern classes will always refer to $B$).
Let ${\cal M}$ be a line bundle over $B$ of $c_1({\cal M})=\eta$, $C$ the 
locus ${s=0}$ for the section\footnote{
the last term is $x^{(n-3)/2}y$ for n odd;
actually $n\equiv 0(2)$ and $\eta\equiv c_1(2)$ was assumed in 
[\ref{FMW}];
$a_r \in \Gamma(B, {\cal M}\otimes {\cal L}^{-r})$}
$s=a_0 z^n+a_2 z^{n-2}x+a_3 z^{n-3}y+ \dots + a_n x^{n/2}$ 
of ${\cal O}(\sigma)^n \otimes {\cal M}$
and ${\cal N}$ a 
('twisting') line bundle over $C$ so that 
\beqa
V=\pi_{2*}({\cal N}\otimes {\cal P}_B)
\eeqa
(where the (relative) Poincare line bundle is restricted to $C\times_B Z$,
${\cal N}$ understood as being pulled back by $\pi_1$); then
\beqa
c_2(V)=\sigma \eta +\omega
\eeqa
where $\omega \in H^4(B)$, i.e. $\omega=c_2(V|_B)$ and $\eta=\pi _*c_2V$.

Because of the triviality of ${\cal P}$ when restricted to $\sigma$ one
gets using Grothendieck-Riemann-Roch for 
$\pi:C\times_B \sigma(B)=C\rightarrow B$ that 
\beqa
\pi_*(e^{c_1({\cal N})} Td(C))=ch(V) Td(B)
\eeqa
With the condition $c_1(V)=0$ one finds
\beqa
c_1({\cal N})=-\frac{1}{2}(c_1(C)-\pi^* c_1(B))+\gamma
\eeqa
where $\gamma \in H^{1,1}(C,{\bf Z})$ with 
$\pi_* \gamma=0 \in H^{1,1}(B,{\bf Z})$ 
(actually $\gamma$ can be half-integral).
One has then
\beqa
\omega=-[\frac{n^3-n}{6}\frac{c_1^2}{4}+\frac{n}{8}\eta(\eta-nc_1)+
\frac{1}{2}\pi_*(\gamma^2)]
\eeqa

Actually the last two terms combine: the only
general elements of $H^{1,1}(C,{\bf Z})$ are 
$\sigma|_C$ and $\pi^* \beta$ (for $\beta \in H^{1,1}(B,{\bf Z})$), 
which have because of $C=n\sigma + \pi ^*\eta$
the relation $\pi_* (\sigma|_C)=\pi_* \sigma(n\sigma+\pi^*\eta)=
\pi_* \sigma(-nc_1+\pi^*\eta)=\eta-nc_1$;
so $\gamma=\lambda (n\sigma-\pi^*(\eta-nc_1))$ (with $\lambda$ 
half-integral if $(K_C^{-1}\otimes K_B)^{1/2}$
does not exist as a line bundle) and 
$\pi_*(\gamma^2)=-\lambda^2 n\eta(\eta-nc_1)$.
So for the generator $\gamma_0$, say, corresponding to 
$\lambda=1/2$, the term completely disappears leaving 
the $\eta$-independent piece
\beqa
\omega (V_{\gamma_0})=-\frac{n^3-n}{6}\frac{c_1^2}{4}
\eeqa
\\ \\
{\bf 1.2 Bundle moduli and F-theory moduli for $G=SU(n),E_8$}\\ \\
$\underline{SU(n)}$

In [\ref{FMW}] a natural map (for $G$ general) 
was given between the (even)
bundle parameters and the $F$-theory parameters (as far as complex
structure is concerned). Their number on the heterotic side 
was also checked by an index 
computation. Of more immediate concern for our purposes is the 
expression of this number in terms of the spectral surface.

Let us first recall that the moduli space ${\cal M}$ of the
bundle is fibered ${\cal M}\rightarrow {\cal Y}$ with
fibre $Jac(C)$. This picture emerges if you consider the Leray spectral
sequence for $\pi :Z\rightarrow B$, which gives for the first-order
deformations\footnote{As our base $B$ will always be
rational and
$C$ will be generically smooth the first-order deformations will
actually be unobstructed [\ref{FMW3}].} $H^1(Z,adV)$
\beqa
0\rightarrow H^1(B,R^0\pi_*adV)\rightarrow H^1(Z,adV)
\ra H^0(B,R^1\pi_*adV)\ra 0
\eeqa
where the first term shows the tangent space to the fibre whereas 
the last term exhibits the space whose projectivization gives the 
global sections of the global moduli object ${\cal W} \ra B$ (cf. 
[\ref{FMW}] and section 2). 

Now we will consider $n_e-n_o$, the 
difference of bundle moduli even respectively odd under the involution $\tau$.
We will establish the relation
\beqa
n_e-n_o=h^{2,0}(C)-h^{1,0}(C)
\eeqa

Before we come to it a number of remarks are in order. First, 
just as in 8D the spectral points in the elliptic curve represent the 
degrees of freedom of the bundle, one should expect by the 
principle of adiabatic extension that the deformations 
(whose number is $h^{d,0}(C)$ for a d dimensional spectral object, 
as its normal bundle equals its canonical bundle) 
of the spectral
object in its Calabi-Yau will represent the even deformations.
So it came actually out 
in the 6D case where the quaternionic dimension of the 
vector bundle moduli space was identified [\ref{BJPS}] with the genus
of the spectral curve as 
in 6D the relation $C=n\sigma + \pi ^*\eta$ means $C=n\sigma+kE$ 
with\footnote{note that 
$k-2n=\sigma \cdot C=h^0(B,\pi_*V)
=h^0(B,R^1\pi_*V)=h^1(K3,V)=-\chi (K3,V)=c_2(V)-2n$}
$k=c_2(V)\in {\bf Z}$ and $E$ the elliptic fibre of $Z=K3$; so
$K_C=C^2=2nc_2(V)-2n^2$ giving the result for $g_C$.
The authors of [\ref{BJPS}] went on to a 4D compactification. 
In a case by case analysis, for 
$B$ a Hirzebruch surface, they matched  
the number of holomorphic two-forms on the
spectral surface $C$ with the corresponding relevant part of the complex 
deformations of the corresponding F-theory four-fold $X^4$. 
They did this by counting 
monomials preserving the type of singularity corresponding to 
the unbroken gauge group.
(This, like the computations in [\ref{FMW}], concerns  a case 
without odd moduli corresponding to $H^{1,0}(C)$
respectively $H^{2,1}(X^4)$). Now one computes with
$C=n\sigma+\pi^* \eta, c_2(Z)=(c_2+c_1^2)+
10c_1^2+12\sigma c_1, \sigma ^2=-\sigma c_1,
\pi_*\pi^*=id,\pi_*\sigma=1$ and Noethers theorem that
\beqa
1+h^{2,0}(C)-h^{1,0}(C)&=&
\frac{c_2(C)+c_1^2(C)}{12}|_C=\frac{c_2(Z)C+2C^3}{12}\nonumber\\
 &=&n+\frac{n^3-n}{6}c_1^2+\frac{n}{2}\eta (\eta-nc_1)+\eta c_1
\eeqa
Now as the ordinary index $\chi (Z, adV)$ vanishes by Serre duality
one has to consider the $\tau$-equivariant
index 
$\chi_{\tau}(Z, adV)=\sum_{i=0}^3(-1)^iTr_{H^i(Z,adV)}\tau$. Using
the projector $\frac{1+\tau}{2}$ one computes for 
\beqa
I=-\frac{1}{2}\chi_{\tau}(Z,adV)
=-\sum_{i=0}^3(-1)^iTr_{H^i(Z,adV)}\frac{1+\tau}{2}
=n_e-n_o
\eeqa
(note that Serre duality interchanges now the even and odd subspaces 
of the respective cohomologies) that [\ref{FMW}]
\beqa
I=rk-\sum_j\int_{U_j}c_2(V)=n-1-4\omega_{\gamma =0}+\eta c_1
\eeqa
(the $U_j$ denote 
the two fixed point sets).\\ \\
$\underline{E_8}$

One can match [\ref{C}] the number of moduli  with the $F$-theory
side, including a number $n_o$ of odd moduli, provided we use
an identification $n_o=h^{2,1}(X^4)$ for the odd moduli.\\
One has for the
heterotic base deformations (assuming that $Z$ has a smooth Weierstrass
model which is general [\ref{Gr}], i.e
has only one section, so that
$h^{1,1}(Z)=h^{1,1}(B)+1=c_2-1=11-c_1^2$)
\beqa
h^{2,1}(Z)=h^{1,1}(Z)-\frac{\chi(Z)}{2}=11+29c_1^2
\eeqa
as the smooth $Z$ has $\chi(Z)=-60c_1^2$ [\ref{KLRY}].
Furthermore one counts the moduli $h^1(Z,adV)=I+2n_o$
of the bundle $V$ by applying [\ref{ACL}] an index computation 
first used in this specific form in [\ref{FMW}] with 
\beqa
I=16+332c_1^2+120t^2
\eeqa
So that in total
\beqa
1+h^{2,1}(Z)+h^1(Z,adV)=28+361c_1^2+120t^2+2n_o
\eeqa
For the $F$-theory four-fold $X^4$ one finds with $\frac{\chi}{6}-8=
h^{1,1}-h^{2,1}+h^{3,1}$ [\ref{SVW}] and $h^{1,1}=2+h^{1,1}(B)=12-c_1^2$ 
(reflecting that there is no unbroken gauge group) the final matching
\beqa
h^{3,1}+h^{2,1}=\frac{\chi}{6}-20+c_1^2+2h^{2,1}=
28+361c_1^2+120t^2+2h^{2,1}
\eeqa
using $c_1^3(B^3)=6c_1^2+2t^2$ [\ref{FMW}] and 
$\frac{\chi}{24}=12+15c_1^3(B^3)$ [\ref{SVW}].\\ \\
{\bf 1.3 Brane impurities}\\ \\
$\underline{E_8}$

For $G=E_8$ the number $n_5$ of non-perturbative heterotic 
fivebranes occurring because of anomaly cancellation is
\beqa
n_5=c_2(Z)-(c_2(V_1)+c_2(V_2))=12+10c_1^2-(\omega_1+\omega_2)
\eeqa
using $c_2(Z)=\eta _Z \sigma+\omega_Z$ with $\omega_Z=12+10c_1^2$ 
and $\eta _Z=12c_1=\eta_1+\eta_2$
in analogy to the decomposition of $c_2(V)$
(remember $\eta_{1/2}=6c_1\pm t$).
Note that the $\sigma$-carrying parts like $\sigma \eta$ are adjusted 
to cancel, as in the 6D story, so the remaining pure $H^4(B)$ parts 
carry the relevant information.
Now $n_5$ was computed [\ref{FMW}] using the relation
\beqa
\omega_{1/2}=-40c_1^2-15t^2\mp45tc_1
\eeqa
to be
\beqa
n_5=12+90c_1^2+30t^2
\eeqa
This was matched, {\it in case} $\gamma=0$ and no moduli are 
odd under the fibre involution, with the number of 
space-time filling three-branes 
$n_3=\frac{\chi(X^4)}{24}=12+90c_1^2+30t^2$ {\it in case} 
of zero four-flux $G=\frac{dC}{2\pi}$ of a dual F-theory compactification
on the elliptically fibered four-fold $X^4\rightarrow B^3$.\\ \\
$\underline{SU(n)}$

The foregoing has the following generalization to
$G=SU(n)$ (cf. [\ref{AC}]). Now one is working for the sake of 
computation of cohomological data (not for doing $F$-theory on it)
on the fiberwise resolved four-fold.\\
Using this device one computes now $h^{1,1}=2+h^{1,1}(B)+16-rk$ 
which leads with the index formula [\ref{FMW}]
\beqa
I-rk=-4(c_2V-\eta\sigma)+\eta c_1=-4\omega_Z+4n_5+12c_1^2
\eeqa
and $h^{2,1}=n_o$ and
\beqa
h^{3,1}=h^{2,1}(Z)+I+n_o+1=12+29c_1^2+I+h^{2,1}
\eeqa 
to (using $\frac{\chi}{6}-8=h^{1,1}-h^{2,1}+h^{3,1}$ [\ref{SVW}])
\beqa
\frac{\chi}{24}=2+\frac{1}{4}(40+28c_1^2+I-rk)=12+10c_1^2-\omega_Z+n_5=n_5
\eeqa

\section{Heuristic considerations: continuous and discrete data}

\setcounter{equation}{0}

A convenient point to start with is a preliminary comparison of the 
discrete data. The most notable features are: \\
${\bf i)}$ shifted integrality (to half-integrality), \\
${\bf ii)}$ restriction to the subspace $ker\; \pi$ respectively 
primitiveness and\\ 
${\bf iii)}$ correction contribution in $n_5$ respectively $n_3$.

ad ${\bf i)}$  let $G_i$, $i=1,2$, be the 
projections associated with the stable degeneration
$X^4\rightarrow X^4_{deg}=W_1 \cup_Z W_2$.
The analogy in the data concerned with the discrete part of the 
twisting degrees of freedom (cf. below) is represented in the 
following juxtaposition: on the heterotic side one has 

\beqa
\gamma=\frac{c_1(C)-\pi ^* c_1(B)}{2}+c_1({\cal N}),
\eeqa
where the last term is an element of integral cohomology whereas the 
square root $(K_C^{-1}\otimes K_B)^{\frac{1}{2}}$ does not necessarily 
exist as a line bundle. Similarly one has on the F-theory side

\beqa
G=\frac{c_2}{2}+\alpha 
\eeqa
where $\alpha \in H^4(X,{\bf Z})$, but $c_2$ is not necessarily even.
Strictly speaking one should consider here the projected 
$G_i$ ($i=1,2$).\\

ad ${\bf ii)}$  the $G$ admissible in an $N=1$ supersymmetric 
compactification are 
in $ker(J\wedge \cdot)$ [\ref{BB}]. 
The last condition comes down 
for the relevant projected classes in $H^{2,2}(W)$ to the following:
on the heterotic side the actual spectral cover construction will
in the $E_8$ case involve the corresponding fibration of $dP_8$ surfaces 
over $B$ (the section of $dP_9$ blown down); now, the embeddings of 
the 8D heterotic elliptic curves in the 8D del Pezzos patch together 
to an embedding of $Z$ in the $W_i$. This gives rise to maps 
$H^*(W)\rightarrow H^*(Z)\rightarrow H^{*-2}(B)$ 
by restriction respectively integration over the fibre; 
but for the $dP_8$ the anticanonical class given by the 
elliptic curve $E$ is ample, so actually the $ker\, (J\wedge \cdot)$
condition is reflected in a $ker (H^{2,2}(W)\rightarrow H^{2,2}(Z))$
condition, respectively, if one combines with the integration 
over the fibre,
in a $ker(H^*(W)\rightarrow H^*(Z)\rightarrow H^{*-2}(B))$ condition;
one has then to divide 
out the class dual to $S_b$, the del Pezzo fibre of $pr:W\rightarrow B$,
corresponding to a differential form supported on the base,
which is mapped to zero in the integration over the $\pi:Z\rightarrow B$
fibre. So finally the space we are concerned with is the 
$(ker:W\rightarrow B)/S_b {\bf Z}$ part in $H^{2,2}$ (cf. 
the second diagram below). So the primitiveness condition is the analogue
of the condition 
$ker \pi:H^{1,1}(C,{\bf Z})\rightarrow H^{1,1}(B,{\bf Z})$ on $\gamma$.

ad ${\bf iii)}$  note  that a
typical value for $G$ such as $\frac{1}{2}c_2(X^4)$ gives non negative 
$G^2$ whereas $\gamma^2$  is negative by the Hodge index theorem. So, 
comparing the heterotic contribution of $\gamma^2$ in eq. (1.5)
\beqa
n_5(\gamma)=n_5(\gamma=0)+\frac{1}{2}\pi_*(\gamma^2)
\eeqa

with the formula [\ref{DM}]
\beqa
n_3=\frac{\chi(X^4)}{24}-\frac{1}{2}G^2\; ,
\eeqa
we are led to expect an association letting $\gamma_i$ correspond 
with $G_i$ giving
\beqa \label {-1}
\pi^i_*(\gamma_i^2)=-G_i^2.
\eeqa
This would fit in and actually
complete the general scheme of a duality dictionary 
beyond the previously considered cases of relating 
$h^{2,0}(C)$ and $h^{3,1}(X^4)$ respectively elements of $H^{1,0}(C)$
and $H^{2,1}(X^4)$ in a satisfying way 
(cf. [\ref{BJPS}],[\ref{FMW}], section 1 
and the introduction).
Together with the proposed identification
of the discrete moduli one gets a dictionary of elements related by 
a $(1,1)$ Hodge shift 

\begin{center}
\begin{tabular}{c|c} 
$C$ & $X^4$ \\ \hline
$H^{2,0}$ & $H^{3,1}$ \\
$H^{1,0}$ & $H^{2,1}$ \\
$H^{1,1}$ & $H^{2,2}$ \\
\end{tabular}
\end{center}
where 
in the first line the deformations of $X^4$ preserving the given type 
of singularity (corresponding with the unbroken gauge group; actually
we will consider the parts in the $W_i$) are understood,
in the second line a part of the relative jacobian (see below) 
is understood,
and in the last line the subspaces $ker\, \pi_*$ respectively 
$ker\, (J\wedge \cdot)$. 

A naive way to obtain the association of $\gamma_i$ with $G_i$ is via 
the cylinder map [\ref{K1}], which we describe below (\ref{cyl}). This  
replaces each point in $C$ by a (complex projective) line $L$ lying above 
it in the del Pezzo. Indeed, $L^2=-1$, suggesting the desired relation 
(\ref{-1}). Unfortunately, this {\em fails}, as the right hand side of (\ref{-1}) 
gets contributions also from distinct lines which intersect in the del Pezzo. 
The full story is a bit more subtle. We will see that $H^i(C)$ breaks into 
several isotypic pieces (five of them, for $E_8$). The values of $\gamma$ 
coming from bundles all live in one of these isotypic pieces, the distinguished 
piece. On this piece the cylinder map changes the intersection numbers not 
by $-1$ but by a factor of $-60$ (for $E_8$). Furthermore, we will see 
(in section (3.2)) that the 
cylinder map itself is divisible (by $60$) on this distinguished piece, so the 
correct association sends $\gamma$ to $\frac{1}{60}$ times its cylinder. 

We will actually have 
to insert an intermediate step in relating the data from the spectral
surface $C_i$ ($i=1,2$), corresponding to the bundle $V_i$, to $W_i$. For
this note first that the (even) deformations of $V_i$ correspond to those
deformations of $W_i$ which preserve fiberwise the elliptic curve $E$
common with the heterotic side, so preserving in total 
the Calabi-Yau $Z$ common to the $W_i$: their number is given by 
the dimension of 
$H^1(W_i,T_{W_i}\otimes {\cal O}(-Z))\cong H^{3,1}(W_i)$.
These are the deformations
in $H^{3,1}(X^4)$ which are relevant to the respective bundle. Second,
under the stable degeneration $J^3(X)$ splits off the abelian
varieties $J^3(W_i)$, which contain the pieces relevant for the 
comparison. Third this construction interprets {\it those} elements
of $H^{2,2}_{prim}(X^4,{\bf Z})$ that are captured by the corresponding 
parts in the $W_i$ cohomology (for the relation of the primitiveness
condition to the $W,Z$ geometry cf. above).

To include also the continuous part in the picture we have to discuss
more fully the issue of the twisting data which can
occur in piecing together bundles over $Z$ if we have a description 
for bundles over an elliptic curve $E$. In this paragraph we 
will review the description of the twisting data based on spectral covers, 
as given in  [\ref{FMW2}]. In the remainder of the paper we will work 
instead with the description given in [\ref{D}], based on the cameral cover.
So let  ${\cal M}_E$ be
the moduli space of semistable $G$-bundles over $E$, respectively
${\cal M}_{Z/B}$ the relative object (for $G=E_8$ we will 
not allow cuspidal fibers) and finally  $\Xi$ the 
(locally existing) universal bundle over
$E\times {\cal M}_E^0$ respectively over $Z\times _B {\cal M}_{Z/B}^0$ (the 
superscript $0$ denoting the smooth locus of the moduli spaces, where then
such a universal object exists locally. Now a $G$ bundle over $Z$
which is fiberwise (over the open subset of $B$ over which lie 
smooth fibers) semistable gives a section (over that subset) 
of ${\cal M}_{Z/B}$. Our aim is 
to describe conversely the (possibly obstructed) existence and 
(non)-uniqueness of a bundle given such a section. The idea is of course 
, as far as possible, to pull-back a universal bundle. Now consider 
first the situation in a fibre: to $\Xi$ there is an associated abelian
group scheme of automorphism groups $\underline{Aut} (\Xi)$ over 
${\cal M}_E^0$ (of associated sheaf of sections, say, ${\cal A}$). The
set of possible universal bundles over $E\times {\cal M}_E^0$ 
(if the obstruction in $H^2({\cal M}_E^0,{\cal A})$ 
vanishes) is then 
rotated through under the elements of $H^1({\cal M}_E^0,{\cal A})$. In 
the relative version where a section $s$ of 
${\cal M}_{Z/B}^0 \rightarrow B$ is given the relevant space is
correspondingly $H^1(B,{\cal A}_B(s))$.

To make this explicit, let us start  with $G=SU(n)$ and
display the decomposition of the twisting data $H^1({\cal A}_B)$ 
into the continuous part (the relative jacobian $J(C/B)$)
and the discrete part (the multiples of the $\gamma$ class).
Both groups are taken up to possible finite group discrepancies. 
Additionally, in the last line we must allow half-integral cohomology.

\begin{center}
\begin{tabular}{ccccccccc} 
 & & $0$ & & & & & & \\
 & & $\downarrow$ & & & & & & \\ 
$0$ & $\rightarrow$ & $J(C/B)$ & $\rightarrow$ & $Pic_0 \, C$ & 
$\rightarrow$ & $Pic_0 \, B$ & $\rightarrow$ & $0$ \\
 & & $\downarrow$ & & $\downarrow$ & & $\downarrow$ & &  \\
$0$ & $\rightarrow$ & $H^1({\cal A}_B)$ & $\rightarrow$ & $Pic \, C$ & 
$\rightarrow$ & $Pic \, B$ & $\rightarrow$ & $0$ \\
 & & $\downarrow$ & & $\downarrow$ & & $\downarrow$ & &  \\
$0$ & $\rightarrow$ & $\gamma {\bf Z}$ & $\rightarrow$ & 
$H^{1,1}(C,{\bf Z})$ &  $\rightarrow$ & $H^{1,1}(B,{\bf Z})$ & 
$\rightarrow$
& $0$ \\
 & & $\downarrow$ & & & & & & \\
 & & $0$ & & & & & & \\
\end{tabular}
\end{center}
Now, when one tries to transfer these results to the 
(D- and especially to the) E-series, one faces at first the following 
problem. For the E-series one does not describe [\ref{FMW}] the bundles
via the spectral cover construction but instead via the associated 
del Pezzo fibration, giving not a covering of $B$ but a fibration over
it by surfaces. This is related to the following 
crucial fact (cf. [\ref{KMV}]): consider the type IIA string on an
elliptic K3 with ADE singularity times $T^2$; the $N=1$ content of this
4D $N=4$ theory includes three adjoint chiral fields $X$, $Y$, $Z$ (with 
superpotential Tr[[$X$,$Y$],$Z$]), whose Cartan vevs (Higgs branch)
correspond to blowing up respectively deforming the singularity respectively giving
Wilson lines to the ADE gauge group on $T^2$; the R-symmetry
induces an equivalence of the corresponding moduli spaces. 
This gives the main theorem on the structure of the moduli space 
${\cal M}_G$ of flat $G$-bundles on an elliptic curve (cf. [\ref{FMW}],
and also [\ref{G}],[\ref{CCML}],[\ref{HM}] for partial identifications
of the relevant mirror map in this connection).

Concretely
let us take as elliptic curve $E=P_{1,2,3}[6]$ of equation
$e:=z^6+x^3+y^2+\mu zxy=0$ 
leading (with $s$ of sect. 1.1) 
to the deformation {$e+vs$} of the $SU(n)$ singularity
showing at the same time Looijenga's moduli space  
${(a_0, a_2, a_3,\dots,a_n)\in {\bf P}^{n-1}}$ of flat $SU(n)$ bundles
over $E$ as well as the 0D spectral geometry consisting of n points 
${(e=0)}\cap {(s=0)}$ on $E$. 
Note that {\it in this case of the $A_n$ group} it is possible to effectively
replace a 2D geometry of ${\bf P}^1$'s by the zero dimensional 
representatives as $v$ occurs only linear and so in the process of 
period integral evaluation to describe the variation of Hodge structure 
relevant here can be integrated out. For the general phenomenon
relating even (0D to 2D, of symmetric intersection form) or odd (1D to 3D,
of antisymmetric intersection form) cohomology cf. [\ref{L}]; the same 
relation underlies the extraction [\ref{KLMVW}]
of the 1D Seiberg-Witten curve from
the 3D periods of a Calabi-Yau and the relation between $K3$ 
singularities and gauge groups $ADE$.

By contrast the same decoupling phenomenon
{\it does not} take place for the the $E_k$ case: there one finds instead 
for the deformation 
$e+\sum_{i=1}^6 a_iv^iz^{6-i}+b_2v^2x^2+b_3v^3y+b_4v^4x$ of zero locus
$dP_8={\bf P}_{1,1,2,3}[6]$ showing the 2D
spectral geometry of the del Pezzo surface with 
$H^{1,1}(dP,{\bf Z})^{\bot _E}=E_8$ and moduli space 
${\bf P}_{1,2,3,4,5,6,2,3,4}$.(Correspondingly there occurs a situation
involving the $E$ groups, where the Coulomb branch of an $N=2$ system
does not reduce to a Riemann surface but a description in terms of 3-form
periods has to be given [\ref{KMV}].)

It is interesting to note that this phenomenon admits also a 
representation-theoretic explanation. The Weyl group of $A_n$
admits a small permutation representation, the (n+1)-dimensional one, 
which decomposes into the sum of only two irreducible representations:
the trivial one and the weights, ${\bf Z}[W/W_0]\cong {\bf 1}\oplus 
\Lambda $. By contrast every permutation representation of $W_{E_n}$ 
contains at least three irreducible constituents. This means that 
the cohomology of every
associated spectral cover will contain some additional pieces.
To get an object with the right cohomology, i.e. the distinguished 
isotypic piece mentioned above, one must either go up in
dimension, or restrict attention to classes which transform correctly 
under some correspondences.  

Note that the the effective replacing of the ${\bf P}^1$ classes by
points accounts for the missing dimensions causing the mentioned (1,1)
shift in cohomology when comparing the dual results. Namely
the description in the $E_k$ case is already well adapted to the
F-theory picture of having a fibration $W\rightarrow B$ (for each
bundle) of del Pezzo surfaces over $B$. As mentioned 
the 8D heterotic
elliptic curve is contained in the both 8D del Pezzo, so is the $Z$ in 
the $W_i$ giving rise to maps $H^*(W)\rightarrow H^*(Z)\rightarrow 
H^{*-2}(B)$ and to the diagram involving now the intermediate jacobian
$J^3$ (cf [\ref{FMW2}] for the analogous situation in 6D) and interpreting
$H^1({\cal A}_B)$ as (relative) Deligne cohomology\footnote{
compare the long exact sequence in Deligne cohomology
\beqa
H^3(\cdot ,{\bf Z})\rightarrow H^3(\cdot ,{\bf C})/F^2 H^3(\cdot ,{\bf C})
\rightarrow H^4_{{\cal D}}(\cdot )\rightarrow H^4(\cdot ,{\bf Z})
\rightarrow H^4(\cdot ,{\bf C})/F^2 H^4(\cdot ,{\bf C})
\eeqa
}
(which relates
to $J^3$ as $Pic$ to $Pic_0$)

\begin{center}
\begin{tabular}{ccccccccc} 
 & & $0$ & & & & & & \\
 & & $\downarrow$ & & & & & & \\ 
$0$ & $\rightarrow$ & $J^3(W/B)$ & $\rightarrow$ & $J^3(W)$ & 
$\rightarrow$ & $J(B)$ & $\rightarrow$ & $0$ \\
 & & $\downarrow$ & &  & & & &  \\
 & & $H^1({\cal A}_B)$ & &  &  &  &  &  \\
 & & $\downarrow$ & & & & & &  \\
$0$ & $\rightarrow$ & $H^{2,2}_0(W,{\bf Z})$ & $\rightarrow$ & 
$H^{2,2}(W,{\bf Z})/S{\bf Z}$ &  $\rightarrow$ & $H^{1,1}(B,{\bf Z})$ & 
$\rightarrow$ & $0$ \\
 & & $\downarrow$ & & & & & & \\
 & & $0$ & & & & & & \\
\end{tabular}
\end{center}

To actually establish this picture we will proceed in three steps:\\ \\
I) first we describe a generalized spectral cover construction 
for arbitrary group $G$ (or in particular, an ADE group.) 
The main point one has to take into account is, that
for $G\neq SU(n)$ the fibre of the spectral covering is of a more
'entangled' nature as the covering group is no longer the 
symmetric group.\\ \\
II) Then we make use of the connection between ADE root systems and 
del Pezzo surfaces (cf. for example [\ref{MS}],[\ref{DKV}],[\ref{Ma}],
[\ref{Dem}]), the
root system describing a certain part of the $H^{1,1}(dP,{\bf Z})$
(denoting the obvious intersection), so that the variation of the fibre
of the spectral cover over $B$ describes the variation of
certain (-1) curves $l$ in their variation in a family of surfaces 
over $B$
(expressing the effective replacement of these lines by points, causing
the (1,1)-shift). This leads also to the necessary  relation between
$G_i^2$ and $l^2 \pi^i_*\gamma ^2=-\pi^i_*\gamma ^2$.\\ \\
III) Thirdly we make the transition to the F-theory side,
which for example in the case $G=E_8$ consists just in relating the
$dP_8$ fibre from the heterotic side to the $dP_9$ fibre from the
F-theory side (blowing down a section brings one back to $dP_8$). 
In general this comes down [\ref{FMW}] 
to see that the occurrence of a section $\theta:B\rightarrow X$
of $H$-singularities ($H$ the commutant of $G$ in $E_8$)
causes the split off from $J^3(X)$
of a certain factor involving cycles of special behavior along 
$\theta$. Remember that 
under the stable degeneration $J^3(X)$ splits off the abelian
varieties $J^3(W_i)$, which contain the pieces relevant for the 
comparison.\\
We take along
also the necessary extension of the picture by inclusion of the 
discrete data.\\
Before entering the general description let us give a short account of
step I) in the framework described already above.
Actually we will see 
the spectral cover as parametrization of exceptional lines
in a surface fibration over $B$.
This occurs by taking into account the description of the 
'enlarged' root system in surface cohomology (cf. the appendix). Note 
that as the same moduli space ${\cal W}_G$
parametrizes $G$ bundles over an elliptic
curve $E$ and del Pezzo surfaces $dP_G$ (with $E=-K$ fixed) one gets
by adiabatic extension over the base $B$ that to the bundle $V$ over $Z$
corresponds a fibration $W^{het}_G \ra B$ of $dP_G$ surfaces via
pulling back the universal object (now the universal surface not the 
universal bundle) along the section 
$s:B\ra {\cal W}_G={\cal M}_{Z/B}$.

Let us conclude this section  with a description of
the spectral cover construction for $G$ an ADE group.
The covering $C\rightarrow B$ will be (cf. [\ref{DD}],[\ref{FMW2}])
locally modelled on (pulled back from)
the covering of degree $d=\frac{|W|}{|W_0|}$\beqa 
{\cal T}&=&(E\otimes \Lambda _c)/W_0\nonumber\\
\downarrow& & \\
{\cal M}&=&(E\otimes \Lambda _c)/W\nonumber
\eeqa
where $W_0\subset W$ is the stabilizer of an element 
$\lambda \in {\bf h}$ whose Weyl orbit spans ${\bf h}$ over 
${\bf C}$;
for example\footnote{
${\bf h}$ the Cartan Lie algebra, $R$ the set of roots, $d=|R|$,
$Q(R)$ the lattice generated by them, $\Lambda _c$ the coroot lattice 
in the Cartan (denoted $\Lambda$ in [\ref{FMW2}]; 
$\tilde{\alpha}$ the quasi-miniscule, non-miniscule weight}
$\lambda$ could be (the dual of) the maximal root 
$\tilde{\alpha}$ with 
$W\lambda=R$.\\

The evaluation on $\lambda$ gives the epimorphisms 
${\bf C}[W/W_0]\rightarrow {\bf h}$ and 
\beqa
({\cal O}_{E\otimes \Lambda _c}\otimes _{\bf C}{\bf C}[W/W_0])^W
\rightarrow ({\cal O}_{E\otimes \Lambda _c}\otimes _{\bf C} {\bf h})^W
\eeqa 
where the sheaf of W-invariant sections of the trivial vector bundle
${\cal O}_{E\otimes \Lambda _c}\otimes _{\bf C} {\bf h}$ is coherent
on ${\cal M}=(E\otimes \Lambda)/W$ and locally free over ${\cal M}^0$,
being equal there to the {\it cotangent bundle}\footnote{
occurring in the
exponential sequence 
$0\rightarrow \underline{\Lambda}\rightarrow Lie {\cal A}\rightarrow 
{\cal A}^0\rightarrow0$, where ${\cal A}={\cal A}^0$ for example on the
(Zariski) open subset of split bundles, where ${\cal A}={\bf C}^{*r}$; 
also ${\cal A}_B(s)={\cal A}^0_B(s)$ for a generic section and $G\neq
A_1, A_2$ (and no cuspidal fibres for $G=E_8$)}
$\Omega ^1_{{\cal M}^0}=Lie {\cal A}$. The ensuing map
$Lie J(C)=H^1(C,{\cal O}_{{\cal T}})\rightarrow 
H^1(Lie {\cal A})=Lie H^1({\cal A})$ presents the continuous part of 
$H^1({\cal A})$ as quotient of the Jacobian $J(C)$, the {\it Prym-Tyurin
variety} of the spectral cover. For $G=SU(n)$ respectively
$SO(2n)$ it reduces to the ordinary Jacobian\footnote{as the 
identification $W_{A_r}\cong S_{r+1}$ shows a certain 'disentangledness'
of the fibre structure in that case,
in contrast to the cases where $W$ is only a subgroup of a symmetric
group, showing that the fibre elements are not on equal footing} 
respectively Prym variety (for the general construction cf. Appendix 
and [\ref{K2}]).\\ \\

\section{Bundle moduli via cover constructions}

There are three geometric objects, each of which can be used 
to encode the even moduli of the heterotic theory. These are the 
spectral cover\footnote{As we will from now on have to distinguish two
'covers' associated to $B$ we switch to the notation $\ov{B}$ for 
the spectral cover (previously denoted $C$),  
and $\widetilde{B}$ for the cameral cover.}  
, the cameral cover, and the del Pezzo fibration. We will 
first recall the description of these objects 
and show that they represent equivalent data.
We will then proceed to the identification of the relevant parts of their 
cohomologies. We fix throughout the heterotic space $Z$, which is 
fibered over a 
base manifold $B$, with elliptic fibers which we assume are 
in Weierstrass form.  Although the bases $B$ which arise in actual 
heterotic compactifications 
are severely restricted, we will not need to make any such assumptions 
about $B$. \\

\noindent{\bf 3.1 The three ``covers"}\ \

{\underline{\bf  Spectral covers}}

The first and most familiar object is the spectral cover 
$\ov{\pi}: \ov{B} \ra B$ which was already mentioned 
in the previous section. For any subgroup $W_0 \subset W$, we can consider a 
spectral cover which is locally modelled on 
the covering $(Hom (\Lambda,E))/W_0\ra (Hom (\Lambda,E))/W$,
of degree $d=\frac{|W|}{|W_0|}$. Here 
$\Lambda$ is the character lattice of G (dual to the $\Lambda _c$ used 
in the previous section) 
and $W_0\subset W$ is an arbitrary subgroup, for example $W_0$ could be 
the stabilizer of an element 
$\lambda \in {\bf h}$. one important case is the cover of smallest degree, 
corresponding to a minuscule weight $\lambda$. 
For groups $G=A_n,D_n,E_6,E_7,E_8$ , this is a branched 
cover of degree $n,2n,27,56,240$ respectively. The opposite extreme 
is when $W=0$. There are also intermediate possibilities, such as the 
one corresponding to the maximal root 
$\lambda = \tilde{\alpha}$ with $W\lambda=R$.  In the context of $G=E_8$ 
we will refer to the smallest cover simply as ``the" spectral cover, although  
other covers also occur naturally (cf. equation (\ref{L3F}).)

{\underline{\bf  The cameral cover}}

The cameral cover $\widetilde{\pi}: \widetilde{B} \ra B$, 
on the other hand, is an abstract  (unembedded) cover, with an 
action of $W$ which, over nice points of $B$, 
is simply transitive. It parametrizes {\em Weyl  chambers} in a moving family of 
Cartan subalgebras. More precisely,  the cameral cover is modelled  on the 
$W$-cover ${\bf h} \ra {\bf h}/W$.
Equivalently, over the open set 
of nice points of $B$, it is modelled on $G/T \ra G/N$. 
Here $T$ is a maximal torus and $N$  is the normalizer in $G$ of $T$, so 
points of $G/N$ parametrize  the possible 
maximal tori, while a point of $G/T$ includes the additional choice 
of a  Borel subgroup containing $T$, which amounts to the same as 
the choice of a chamber in the 
corresponding Cartan. The extension to all of $B$ is modelled on an appropriate 
partial compactification $\ov{G/T} \ra \ov{G/N}$ constructed in [\ref{D2}].
\footnote{points of $\ov{G/N}$ parametrize regular centralizers in $G$, i.e. 
abelian subgroups whose dimension equals the rank of $G$ and which are the 
commutant of some element of $G$. The maximal tori are regular centralizers.
An example of a regular centralizer which is not a maximal torus is the 
commutant in $G=SL(n)$ of a nilpotent element made up of a single Jordan block.}
We have seen that the even part of the heterotic data amounts to the data 
of a spectral cover; this is the same as specifying the 
cameral cover $\widetilde{B} \ra B$ {\em together with} a $W$-equivariant map 
$v : \Lambda \times \widetilde{B} \ra Z$ 
(or equivalently, a family of maps 
$v_{\lambda} : \widetilde{B} \ra Z$ depending linearly on 
$\lambda \in \Lambda$) commuting with the projection to $B$. Each such 
$v_{\lambda}$ induces in particular a map 
$\pi_{\lambda} : \widetilde{B} \ra \ov{B}$ where $\ov{B}$ is the 
spectral cover 
corresponding to the subgroup $W_0 \subset W$ which fixes $\lambda$.
For $\lambda$ in the interior of a chamber, $\pi_{\lambda}$ can be 
expected to be a birational isomorphism between $\widetilde{B}$ and 
the largest spectral cover, given by  $W_0=0$.
The main difference  between  the 
$W_0=0$ spectral cover and $\widetilde{B}$ is that the former sits, 
fiber by 
fiber, inside $Hom(\Lambda ,E)$, while the latter sees only 
information along the 
base $B$ and is correspondingly an abstract (unembedded) cover. 

{\underline{\bf  Del Pezzo fibrtaions}}
Our third geometric object is not a cover, but a fibration $\pi:U \ra B$ 
whose fibers are complex surfaces. Let us describe how this is 
constructed for $G=E_8$, where the fibers are $E_8$ del Pezzo surfaces, 
obtained from ${\bf P}^2$ by blowing up successively $8$ (distinct or possibly 
infinitesimally near) points. The Picard group of such a surface, or its 
second cohomology  $H^2(dP_8, {\bf Z})$, is a rank-9 
lattice, generated by the class $L$ (pullback of a line in ${\bf P}^2$) and the 
8 exceptional curves $E_i$. It contains the anticanonical class 
$F:=3L-\sum_i E_i$,  with $F^2=1$. Correspondingly, sections of the 
anticanonical system form a pencil of elliptic curves passing through 
a base point $p$. We will normalize the embedding of an elliptic curve into 
its del Pezzo by requiring that the zero point $\sigma$ of the elliptic curve 
map to the base point $p$ of the del Pezzo. The primitive cohomology, or 
the orthogonal complement  $H^2_0(dP_8, {\bf Z})$ of $F$, is 
isomorphic to the $E_8$ weight lattice $\Lambda$, generated by  the classes 
$\lambda_i:= E_i - E_{i+1}, \ i=1,...,7$
and
$\lambda_8:=L-E_1-E_2-E_3$.
In fact the Picard group is the 
direct sum of $\Lambda$ and ${\bf Z}F$. In particular, $L$ can be expressed 
as a linear combination of $F$ and the $\lambda_i$. Explicitly, we find:
\beqa\label{L3F}
L-3F=-(5\lambda_1+10 \lambda_ 2+15\lambda_3+12 \lambda_4+
9  \lambda_5+6 \lambda_6+3 \lambda_7+8 \lambda_ 8)
\eeqa
Now given the Weierstrass elliptic fibration
$\pi_Z: Z \ra B$ with section $\sigma: B \ra Z$ and an $E_8$ bundle on $Z$ whose 
restriction to each fiber is semistable and regularizable 
(in the sense of [\ref{D}]),  we get a cameral cover 
$\widetilde{B} \ra B$, hence a pullback family 
$\pi_{\widetilde{Z}} : \widetilde{Z}  := Z \times_B \widetilde{B} \ra \widetilde{B}$
together with a map
$v: \Lambda \times \widetilde{B} \ra \widetilde{Z} $. 
We assume also that the image of this map 
does not contain any singular points of singular elliptic fibers.
Starting with this data, we will construct a del Pezzo bundle $U \ra B$
as well as an embedding of $Z$ into $U$.

We start by constructing a ${\bf P}^2$-bundle $\widetilde{U}_0$ 
over $\widetilde{B}$ containing $\widetilde{Z}$ as a family of Weierstrass 
cubics. This is obtained as projectivization of the rank-3 vector bundle  
$(\pi_{\widetilde{Z}})_*{\cal L}$ for some line bundle $\cal L$ on 
$\widetilde{Z}$, of degree 3 on each elliptic fiber. The simplest choice, 
${\cal L}_0 := {\widetilde{\pi}}^*({\cal O}_Z (3\sigma))$, does {\em not} work. 
Instead, we need to take 
\beqa\label{calL}
{\cal L}:= {\widetilde{\pi}}^*({\cal O}_Z (2\sigma)) \otimes 
{\cal O}_{\widetilde{Z}} (v((L-3F) \times \widetilde{B})),
\eeqa
where $L-3F$ is given by (\ref{L3F}).
Now over $\widetilde{B}$ we have 8 sections $v_{\lambda_i}$
of  $\widetilde{Z}$, and hence of  $\widetilde{U}_0$, labelled by the basis 
$\lambda_i \in \Lambda$. We will blow them up sequentially in $\widetilde{U}_0$.
The fibers of the 
resulting family $\widetilde{U}' \ra \widetilde{B}$ are almost del Pezzo 
surfaces: they become del Pezzos when line configurations of type ADE 
are blown down to produce ADE singularities. This can be done 
simultaneously for the whole family: replace 
$\widetilde{U}'$ by its image $\widetilde{U} \ra \widetilde{B}$ under a 
sufficiently high multiple of the relative anticanonical bundle. 
Now the action of 
the Weyl group $W$ on $\widetilde{B}$ lifts to $\widetilde{U}$: it acts by 
isomorphisms between the del Pezzo fibers, and by Cremona transformations 
on the original ${\bf P}^2$ fibers. Additionaly, if $w \in W$ 
fixes some $\widetilde{b} \in \widetilde{B}$ then it automatically 
acts as the 
identity on the del Pezzo fiber $\widetilde{U}_{\widetilde{b}}$. 
So $\widetilde{U}$ is the pullback to $\widetilde{B}$ of a well-defined 
del Pezzo fibration 
$U := \widetilde{U}/W$ over $B$.
We also see that these operations do not affect the subvariety $Z$ 
which is 
therefore still embedded in the resulting del Pezzo fibration $U$ 
as a family of elliptic fibers.
(Our choice (\ref{calL}) of the line bundle   ${\cal L}$ is 
essentially the unique choice for which the action lifts and such that the 
zero-section $\sigma$ of $Z$ maps to the base-point section $p$ 
of the del Pezzos.)

We have just seen that the even heterotic data, consisting of  an $E_8$ 
spectral cover (or equivalently of a cameral $\widetilde{B}$ plus maps
 $v: \Lambda \times \widetilde{B} \ra \widetilde{Z}$), determine a del Pezzo 
fibration $\pi:U \ra B$.  Conversely, given $Z$ and $U$ we recover the spectral cover 
$\ov{B} \ra B$: it parametrizes the lines in the moving del pezzo fibers of $U$.
Similarly, a point of the cameral $\widetilde{B}$ corresponds to an 8-tuple of 
disjoint lines $E_i$ in the del Pezzo, or equivalently to an isomorphism (preserving 
the intersection forms) from $\Lambda$ to $H^2_0(dP_8, {\bf Z})$.

The upshot then is that our three types of data: a spectral cover, a del Pezzo 
fibration, and a cameral cover with map $v$, are equivalent to each other when
$G=E_8$. The obvious analogue works for type $E_n$:  a del Pezzo of type 
$E_n$ is the successive blowup of ${\bf P}^2$ at $n \leq 8$ (distinct or possibly 
infinitesimally close) points. The character lattice $\Lambda$ of  $E_n$ is still 
isomorphic, as an abelian group with intersection form, to the primitive 
cohomology group  $H^2_0(dP_n, {\bf Z})$.
For type $A_n$ or $D_n$ we use the fact that the corresponding character 
lattices can be embedded into the $E_n$ lattice as the orthogonal complement 
of an appropriate fundamental weight (corresponding to one of the ends of the 
Dynkin diagram). We will therefore {\em define} a del Pezzo fibration of type 
$A_n$ or $D_n$  to be a del Pezzo fibration $\pi:U\ra B$ of type $E_n$ together 
with a section of the family of $E_n$ lattices  $R^2\pi _* {\bf Z}$ which, in each 
fiber, is in the $W$ orbit of that fundamental weight. (For $A_n$, for example, 
this additional  data consists, in each fiber, of specifying the pullback of a line 
of the original $P^2$ .) This extends the correspondence between cameral covers 
(plus $v$), spectral covers, and del Pezzo fibrations from the $E_n$ case to the 
ADE case. We can therefore refer to either of these three types of data as 
{\em spectral data}. In the sequel we will, however, concentrate on the case
$G=E_8$.\\ \\

\noindent{\bf 3.2 The matching of cohomologies} \ \

Since the heterotic moduli are mapped to the family of spectral data, 
it  is crucial to understand the fiber of this map, or the space of twisting data 
for a given cover. Two descriptions of this fiber are in the literature.
In [\ref{D}]  it was seen that this fiber can be identified
with the space of 
principal $G$-Higgs bundles $(P_B, {\cal C})$ on $B$. 
\footnote{A principal $G$-Higgs bundle on $B$ is a principal $G$-bundle 
$P_B$ together with a family ${\cal C} \subset Ad P_B$ of  regular centralizers 
in the adjoint  bundle $Ad P_B$. Regular centralizers were defined in the 
previous footnote. Actually, what is identified with the space of  principal $G$-Higgs 
bundles is the space of {\em regularized} principal bundles with given spectral 
data. In well-behaved situations, e.g. when the restriction of our bundle to each 
elliptic fiber is already regular, the regularization is unique and so can be omitted.} 
The space of  principal $G$-Higgs bundles has nothing to do with our elliptic 
fibration $Z$: the definition involves only the base $B$. So the space of 
principal G-Higgs bundles with given cameral cover $\widetilde{B}$ should 
be describable in terms of $B, \widetilde{B}$ only. 
Such a description was indeed  
found in [\ref{D2}], Theorem 12: if this space is non-empty, it is a principal 
homogenous space over a certain subgroup of  
$Hom_W (\Lambda, Pic(\widetilde{B}))$. More 
precisely, for compact, connected $B$,  this subgroup is the 
{\em distinguished Prym} of the cameral cover, or the kernel,
$Prym_{\Lambda}(\widetilde{B}) := ker(cl)$,
of the natural homomorphism
\beqa\label{cl}
 cl : Hom_W (\Lambda, Pic(\widetilde{B})) \ra H^2(W,T).
\eeqa
Here as before, $\Lambda$ is the character lattice of $G$, and $Hom_W$ 
is the group of 
$W$-equivariant homomorphisms. This Prym has both a continuous and a 
discrete part. (The name "distinguished Prym" was applied in [\ref{D2}] and [\ref{D}],
somewhat ambiguously, both to this Prym and to its connected component. )
A similar description of the fiber (modulo some equivalences for irregular 
bundles)  is given in [\ref{FMW2}] in terms of the object $H^1({\cal A}_B)$ 
described in the previous subsection. For regular bundles, 
the sheaf ${\cal A}_B$ of automorphisms along the fiber, desribed in  
[\ref{FMW2}], is the same as the 
sheaf $\cal C$ of regular centralizers of [\ref{D2}], [\ref{D}].

Two further versions of the Prym can be constructed from our other 
geometric 
objects: the Prym-Tyurin variety $Prym(\ov{B}/B)$, an abelian 
subvariety of  
$Pic^0(\ov{B})$ whose construction we recall below, and the 
relative intermediate 
Jacobian $J^3(U/B)$. Our purpose here is to compare these three. 
We will find that 
$Prym(\ov{B}/B)$ and $J^3(U/B)$ can be identified (up to a finite group) 
with the continuous part of 
$Prym_{\Lambda}(\widetilde{B})$, and that the discrete part too 
(again, up to a possible discrepancy of  a finite group) 
can be identified 
in terms of cohomology groups of $\ov{B}$ and $U$. 
(We remind the reader that our 
del Pezzo fibration $U$, and hence also its 
intermediate Jacobian $J^3(U/B)$, live 
in the heterotic theory. It is therefore possible to obtain their exact 
relationship to the cameral or spectral covers by purely geometric means.
It is 
the intermediate Jacobian $J^3(X)$, which arises in the F-theory, 
which is related 
to all three of our objects only in the stable degeneration limit.)

We are going to compare the relevant parts of the cohomologies of our 
three geometric objects. It is convenient to discuss the rational
cohomology first, then to pass to integer coefficients, and finally to 
consider the cohomology with ${\bf R/Z}$ coefficients, which gives the 
abelian varieties.\\

{\underline{\bf Rational cohomology of the cameral cover}}

Let us start with the rational cohomology $H^i(\widetilde{B}, {\bf Q})$ 
of the cameral cover, which is a representation of $W$. As such, 
it decomposes into {\em isotypic pieces}:
$$H^i(\widetilde{B}, {\bf Q}) = \bigoplus_{\rho} M^{\rho} \otimes 
H^i_{\rho},$$ 
where $\rho$ runs through the irreducible representations 
of $W$, the multiplicity space  $M^{\rho}$ is just the ${\bf Q}$ 
vector space in which $\rho$ acts, and 
$$H^i_{\rho} :=  Hom_W (\rho, H^i(\widetilde{B}, {\bf Q})).$$ 
We will be concerned mostly with two of 
these pieces, where $\rho$ is either the representation $\Lambda$ of $W$ 
on the weights of $G$, or the trivial representation $\bf 1$.\\

{\underline{\bf Rational cohomology of the spectral cover}}

The Weyl group does not act on the spectral cover 
$\ov{B}=\widetilde{B}/W_0$, nor therefore
on its cohomology. Nevertheless, we can decompose $H^i(\ov{B}, {\bf Q})$ 
into its isotypic pieces (cf. [\ref{DD}]) simply by using one of the 
projections $\pi_{\lambda} : \widetilde{B} \ra \ov{B}$ to embed 
$H^i(\ov{B}, {\bf Q})$ into $H^i(\tilde{B}, {\bf Q})$. We write this as
$$H^i(\ov{B}, {\bf Q}) = \bigoplus_{\rho} {M_{\ov{B}}^\rho} \otimes H^i_{\rho},$$ 
with multiplicity space
$M^{\rho}_{\ov{B}} \subset M^{\rho}$ which is the stabilizer of $W_0$ in 
$M^{\rho}$ . Up to conjugation, 
this is independent of the map $\pi_{\lambda}$ used. 
The two representations $\rho = \Lambda, \bf 1$ have the property that they 
are present in the cohomology of {\em every} spectral cover $\ov{B}$: the 
multiplicity space 
$M^{\rho}_{\ov{B}}$ is always non-zero (cf.  [\ref{D2}]).

For every irreducible representation $\rho$ of $W$ there is a 
correspondence $D_{\rho}$  on ${\ov{B}}$ 
which induces on $H^i(\ov{B})$ (a certain multiple $-q$ of) the projection
to a $\rho$-isotypic piece. A general formula for such correspondences is 
given in section 12 of  [\ref{DD}] in terms of an integral vector 
$v \in M^{\rho}$ which is fixed by $W_0$, and  a $W$-invariant pairing 
on $M^{\rho}$. It was seen there that the multiple is given by:
\beqa\label{dimrep}
-q = {\frac{[W:W_0]}{dim\  \rho}}\vert v \vert ^2.
\eeqa
Essentially the same result 
in the case $\rho=\Lambda$ was obtained earlier by Kanev.
(We would like to point out that the analogous correspondence on 
$\widetilde{B}$  exists, but is much simpler, so it is not needed explicitly: 
it is given in the obvious way as a  combination of the
{\em actions} of elements $w \in W$ with weights given by the 
(${\bf Z}$-valued!) characters $Tr_w(\rho)$. Our point here is that 
even though there is no $W$-action on $\ov{B}$,
the $D_{\rho}$ survive on the quotient covers $\ov{B}$ as a sort of 
{\em replacement} for the $W$-action.)

We work this out explicitly in the case of $G=E_8$ and the standard, degree 240, 
spectral cover  $\ov{B}$ whose points parametrize lines on the 
del Pezzo fibers of $U$ . We will describe explicitly 
some of the correspondences on the
isotypic pieces. Let $D_k$ be the family of line pairs in $\ov{B} \times
\ov{B}$ with intersection number $k$; $D_k$ is non empty only
for $k=-1,\dots, 3$. 
Under the action of the ring generated by the $D_k$, 
we find that  the $W$-module ${\bf Q}[W/W_0]$, 
the local system $\ov{\pi}_{*}{\bf Q}$, 
as well as the cohomology $H^*(\ov{B}, {\bf Q})$ each decomposes 
into five pieces corresponding to representations of dimensions 1, 35,
84, 8, 112; the first three are even, the last two odd under the action of 
the Bertini involution $D_3$. The most useful correspondences 
will be the combinations 
\beqa\label{corrs}
D:= \sum kD_k, J:=\sum D_k, D^{\prime}:=D-J.
\eeqa

The eigenvalues of $D$ are 240 on ${\bf 1}$,  -60 on ${\bf 8}=\Lambda $,
and $0$ on the other pieces; on the other hand,  the 
only nonzero eigenvalue of $J$ is 240 on ${\bf 1}$. It follows that 
$D^{\prime}$ acting on $H^*(\ov{B})$
is $-q={\frac{240}{8}}\vert \widetilde{\alpha} \vert ^2 =-60$ times 
the projection to the $\Lambda $ piece, in accordance with the general 
formula (\ref{dimrep}). 
In particular, the image of $D^{\prime}$ acting on $H^i(\ov{B}, {\bf Q})$
is just  the distinguished piece $H^i_{\Lambda}$, while the image of 
$D$ acting on $H^i(\ov{B}, {\bf Q})$ is 
$H^i_{\Lambda} \oplus H^i_1$.
(The correspondence $i$ used in the description given by Kanev 
[\ref{K1}] , [\ref{K2}]  and recalled in the Appendix below is 
$i=D+D_{-1}$, where $D_{-1}=\Delta $ acts as the identity, so the 
eigenvalues of $i$ are shifted by $1$ from the eigenvalues of $D$. 
Note that $D$ and $D^{\prime}$ differ only on the trivial
piece $H^*(B)$; Kanev can thus work with the correspondence 
$i$ (which is equivalent to working with $D$)  and get away with it, 
because he takes the base
$B$ to be $P^1$ which has no $H^1$. We are in a situation where we 
do care about the possible contribution of the cohomology of $B$, so 
we are forced to work with the "correct" correspondence $D^{\prime}$.)

This description has obvious analogues for the ADE groups. We refer 
the reader to [\ref{K2}] for some of the details.

{\underline{\bf Cohomology of the del Pezzo fibration}}

Elementray topological considerations show that for each $i$ the cohomology 
group $H^{i+2}(U,{\bf Z})$ decomposes as the sum of subgroups:
\beqa\label{decomp}
H^{i+2}(U,{\bf Z}) = H^{i+2}(B,{\bf Z}) \oplus 
H^{i-2}(B,{\bf Z})  \oplus H^{i+2}_p.
\eeqa
We will see below that the "pure" term  $H^{i+2}_p$  decomposes further,  as:  
\beqa\label{subdecomp} 
H^{i+2}_p   = H^{i}(B,{\bf Z}) \oplus H^{i+2}_0(U).
\eeqa
We will refer to the summand $H^{i+2}_0(U)$ as the {\em reduced} 
cohomology of $U$. We will see in the next paragraph that, 
over $\bf Q$,  it can be identified 
with the distinguished piece $H^i_{\Lambda}$ which occurred in the cohomologies 
of  $\ov{B}$ and $\widetilde{B}$.

To describe the decomposition (\ref{decomp}), we use  
the projection $\pi: U \ra B$, the 
section $\sigma : B \ra Z$, the  inclusion $j: Z \ra U$, and 
the projection $\pi_Z : Z \ra B$. In terms of these maps, we can describe the 
first two summands more accurately as:
$$\pi^{*}(H^{i+2}(B,{\bf Z}))    \oplus  j_{*}\sigma_{*}(H^{i-2}(B,{\bf Z})).$$
Let
$\alpha := \pi^* \sigma^*j^*,  \ \ \beta :=j_*\sigma_*\pi_*$
be the projections onto these summands. They satisfy
$\alpha^2=\alpha, \beta^2=\beta, \beta \alpha=0$ but, unfortunately,
$\alpha \beta \neq 0$. Nevertheless, their images do fit into a direct sum 
decomposition (over ${\bf Z}$!). In fact, we can find a set of three orthogonal 
projections, namely $ \alpha(1-\beta), \beta, (1-\alpha)(1-\beta)$.
Their images are our summands $H^{i+2}(B,{\bf Z})$, 
$H^{i-2}(B,{\bf Z})$ and  $H^{i+2}_p.$ (Note that  $ \alpha, \  \alpha(1-\beta)$
have the same image, since $\alpha(x)=\alpha(1-\beta)\alpha(x)$.)

For the comparison with  $H^i_{\Lambda}$ we will use the {\em cylinder map}:
\beqa \label{cyl}
c:=i_* p^* : H^i(\ov{B},{\bf Z}) \ra H^{i+2}(U,{\bf{Z}}),
\eeqa
where $p: {\cal P} \ra \ov{B}$ is the natural $P^1$-bundle, and 
$i: {\cal P} \ra U$  embeds each $P^1$ as a line in the del Pezzo. 
In the opposite direction we have 
$$c^* := p_* i^*: H^{i+2}(U,{\bf{Z}}) \ra H^i(\ov{B},{\bf Z}).$$
In [\ref{K1}] Kanev shows the surjectivity of $c$  (over $\bf Q$) 
when the base $B$ is ${\bf P}^1$.
The generalization we will prove is that, after composition with 
$(1-\alpha)(1-\beta)$, the projected cylinder map
$(1-\alpha)(1-\beta)c=(1-\alpha)c$
induces a surjection of  $H^i(\ov{B},{\bf Q})$ onto 
$ H^{i+2}_p  \otimes {\bf Q}$. Further, this is compatible with the isotypic 
decomposition of  $H^i(\ov{B},{\bf Q})$ into five pieces: the cylinder map 
takes the distinguished piece $H^i_{\Lambda} \subset H^i(\ov{B},{\bf Q})$ 
isomorphically to  $H^{i+2}_0,$ while the trivial piece $H^i_{\bf 1}$ 
goes isomorphically to $H^i(B,{\bf Q})$. In fact, the composition $cc^*$ acts
as multiplication  by $240$ on $H^i(B,{\bf Q})$ and by $-60$ on $H^{i+2}_0,$
while $c^*c$ acts by $240$ on ${\bf 1}$, by $-60$ on $\Lambda$, and by $0$ 
on the three other pieces.
(Concerning the trivial piece $H^i_{\bf 1}$, we point out that there are really 
two distinct ways we could have used to map it into $H^{i+2}(U,{\bf{Z}})$, 
namely via $j_*\pi_{Z}^*$ or via $i_*p^*\ov{\pi}^*$. The difference 
$i_*p^*\ov{\pi}^* - 240 j_*\pi_{Z}^*$ is in the image of $\alpha$.)

{\underline{\bf Local systems}}

All in all, we have three ways to realize the lattice $\Lambda$:

\noindent (1) As the isotypic piece $Hom_W(\Lambda,{\bf Z}[W])$

\noindent (2) As the image $D'{\bf Z}[W/W_0]$ of the corespondence $D'$ 
of (\ref{corrs})

\noindent (3) As the reduced part $H^2_0(S, {\bf Z})$ of the cohomology of a
del Pezzo-8 surface $S$.

These extend to three isomorphic local systems over our base $B$:

\noindent (1) $\Lambda_{\widetilde{B}} := 
Hom_W(\Lambda,\widetilde{\pi}_*{\bf Z}_{\widetilde{B}})$

\noindent (2) $\Lambda_{\ov{B}} := D'  \ov{\pi}_*{\bf Z}_{\ov{B}}   $ 

\noindent (3) $\Lambda_U := (R^2 \pi_*{\bf Z}_U)_0$

(Here by "local system" we mean a locally constant sheaf on the open 
subset of $B$ where the covers are unramified, extended by (inclusion$)_*$
to all of $B$.) Indeed, the identification of (1) and (3) is immediate using the 
self-duality of $\Lambda$.  The isomorphism from (3) to (2) is induced by (the 
restriction to the reduced piece of) $c^*$. The inverse map is (the restriction 
to the $\Lambda$ piece of) $-c/60$, which is defined over the integers. (This is 
equivalent to checking that the image of $D'$ acting on ${\bf Z}[W/W_0]$ is the 
same as the image of $\Lambda$ in ${\bf Z}[W/W_0]$. But the natural surjective map 
${\bf Z}[W/W_0] \ra \Lambda$ equals $60$ times the projection, so the projection
$pr_\Lambda: {\bf Q}[W/W_0] \ra \Lambda \otimes {\bf Q}$ itself takes 
${\bf Z}[W/W_0]$ onto $\frac{1}{60}\Lambda$, and therefore $D'{\bf Z}[W/W_0] =D'\frac{1}{60}\Lambda=\frac{1}{60}D'\Lambda=\frac{1}{60} 60\Lambda=\Lambda$.)
As we promised in the previous section, this provides the justification for the 
matching of brane impurities.

{\underline{\bf Spectral sequences}}

The cohomology of the covers can be computed in terms of local systems on $B$:
$$H^i(\widetilde{B},{\bf Z}) = H^i(B, \widetilde{\pi}_*{\bf Z})$$
$$H^i(\ov{B},{\bf Z}) = H^i(B, \ov{\pi}_*{\bf Z})$$
The cohomology of $U$, on the other hand, requires the complex of sheaves
$R\pi_*{\bf Z}$; in other words, it comes out of the Leray spectral sequence 
for $\pi$. But the projections $\alpha$ and $\beta$ of (\ref{decomp}) act on 
$R\pi_*{\bf Z}$, so each summand in (\ref{decomp}) is again computed as 
the cohomolgy of a local system on $B$.  In particular, 
$ H^{i+2}_p =(1-\alpha)(1-\beta)H^{i+2}(U,{\bf Z})=H^i(B,R^2\pi_*{\bf Z}).$
But this local system is, globally, a direct sum of a trivial piece (coming from 
the anticanonical divisors $F$ in the del Pezzos) and the $\Lambda$ piece:
$R\pi_*{\bf Z} = {\bf Z}F \oplus \Lambda_U $. Therefore, its cohomology 
decomposes as claimed in (\ref{subdecomp}). The reduced cohomology 
$H^{i+2}_0(U)$ can therefore be described as $H^i$ of $B$ with coefficients 
in either of the local systems $\Lambda_{\widetilde{B}}, \Lambda_{\ov{B}} , $ or 
$\Lambda_U.$

Next we want to describe the distinguished pieces of the cohomologies of 
$\widetilde{B}$ and $\ov{B}$:
$$H^i_\Lambda(\widetilde{B}, {\bf Z}) := Hom_W(\Lambda, H^i(\widetilde{B},{\bf Z})), 
\ \ H^i_\Lambda(\ov{B}, {\bf Z}) := D'H^i(\ov{B},{\bf Z}).$$
It is convenient to introduce the two spectral 
sequences computing the $W$-equivariant cohomology 
$H^{i+j}_W := H^{i+j}_W(Hom(\Lambda,\widetilde{\pi}_*{\bf Z}_{\widetilde{B}}))$.
Their $E_2$ terms are, respectively,
$$_1E^{ij} = H^i(W,Hom(\Lambda,H^j(\widetilde{B},{\bf Z})))$$
and
$$_2E^{ij} =H^j(B,H^i(W,Hom(\Lambda,\widetilde{\pi}_*{\bf Z}_{\widetilde{B}}))).$$
From $_2E$ we get edge homomorphisms
$$H^{i+2}_0(U,{\bf Z}) = H^i(B, \Lambda_{\widetilde{B}}) = 
H^i(B,Hom_W(\Lambda,\widetilde{\pi}_*{\bf Z}_{\widetilde{B}})) \ra H^i_W,$$
while $_1E$ gives 
$$H^i_W \ra Hom_W(\Lambda, H^i(\widetilde{B},{\bf Z})).$$
Similarly for $\ov{B}$ we find maps
$$ H^i_\Lambda(\ov{B}, {\bf Z}) = D'H^i(B, \ov{pi} _*{\bf Z}) \ra 
H^i(B, D' \ov{pi} _*{\bf Z}) = H^{i+2}_0(U,{\bf Z})$$
which can also be analyzed via spectral sequences. At any rate, we see 
that the natural map between the distinguished pieces for  $\ov{B}$
and $\widetilde{B}$ factors:
\beqa\label{maps}
H^i_\Lambda(\ov{B}, {\bf Z}) \ra H^{i+2}_0(U,{\bf Z}) \ra H^i_W \ra 
H^i_\Lambda(\widetilde{B}, {\bf Z}) .
\eeqa

In order to go further, we need to compute some group cohomologies.
In general, all the higher cohomologies are finite abelian groups.
A useful observation is that $H^1(W,Hom(\Lambda,{\bf Z}[W]))$ vanishes.
(This is the same as 
$$H^0(W, Hom(\Lambda,{\bf Z}[W]) \otimes ({\bf R}/{\bf Z}))\  / \ \
H^0(W, Hom(\Lambda,{\bf Z}[W]) \otimes ({\bf R})).$$
But since $W$ permutes a ${\bf Z}$-basis of ${\bf Z}[W]$, all 
$W$-invariants in the torus come from $W$-invariants in the vector space.)
This implies that the $_2E^{1j}$ terms vanish, while 
$_2E^{0j} = H^j(B,\Lambda_{\widetilde{B}}) = H^{j+2}_0(U,{\bf Z})$.
In particular we find:
\beqa\label{E2gives}
H^3_0(U, {\bf Z}) = H^1_W, 
\ \ 0 \ra  H^4_0(U, {\bf Z}) \ra  H^2_W \ra 
H^0(B,H^2(W,Hom(\Lambda, \widetilde{\pi}_*{\bf Z}_{\widetilde{B}}))).
\eeqa

In order to simplify $_1E$, we will assume that our group $G$ is of 
adjoint type, e.g. this holds for our main interest, $G=E_8$. This means that
$\Lambda$ is the root lattice, and $\Lambda^*:=Hom(\Lambda, {\bf Z})$ 
is the full weight lattice (of the Langlands dual group, but this is immaterial for 
groups of types ADE). In this case one sees easily that 
$_1E^{i0} = H^i(W, \Lambda^*)=0$ 
for $i=0,1$. By an explicit calculation, we checked this also when $i=2$ for adjoint 
$G$ of type $A_n, D_n$ or $E_8$. The next term, $_1E^{30} =H^3(W, \Lambda^*)$ 
is precisely the right hand side $H^2(W,T)$ of (\ref{cl}). It would be nice to know 
that this vanished, but we have not computed it yet. Our conclusions for $_1E$ 
are:
\beqa\label{H1W}
H^1_W = Hom_W(\Lambda, H^1(\widetilde{B},{\bf Z})),
\eeqa
and, under the assumption that $H^2(W,T) = 0$,
\beqa\label{E1gives}
0 \ra  H^1(W,Hom(\Lambda,H^1(\widetilde{B},{\bf Z})))
\ra  H^2_W  \ra Hom_W(\Lambda, H^2(\widetilde{B},{\bf Z})) \ra \\
\ra H^2(W,Hom(\Lambda, H^1(\widetilde{B},{\bf Z}))) \nonumber
\eeqa

{\underline{\bf The twisting data}}

We are finally ready to prove our main result, the identification of  
the distinguished Prym variety $Prym_{\Lambda}(\widetilde{B})$, 
which precisely parametrizes the 
full twisting  data in  the heterotic theory, with the Deligne cohomology 
of the del Pezzo fibration. As we have already emphasized several 
times before, this is an identification between two objects in the heterotic 
theory. When we go, in the next section,  to an appropriate boundary 
component, the Deligne cohomology of the del Pezzo fibration will be 
reinterpreted as the fiber, or twisting data,  in  the fibration of the F-theory 
moduli. The appropriate boundary component is the locus where the
T-dualities are killed, so that the relation between the heterotic 
and F-theoretic fibers can indeed be expected to be described 
there by classical  geometry.

Our main point here is that the continuous data is actually determined by 
the discrete data which we have already analyzed. The maps between 
the lattices in (\ref{maps}) are, of course, morphisms 
of Hodge structures. Therefore, they induce maps between the Abelian 
varieties made out of those lattices. We recall that the family of twisting 
data for the heterotic theory is given by $Prym_{\Lambda}(\widetilde{B})$, 
defined in (\ref{cl}). Its connected part is the Abelian variety made out of 
the lattice
\beqa\label{cont}
_1E^{01} = Hom_W(\Lambda,H^1(\widetilde{B},{\bf Z})).
\eeqa
We want to compare this with the intermediate Jacobian
$J^3(U/B) := J^3_0(U)$. But this is the Abelian variety made out of the lattice 
$H^3_0(U,{\bf Z})$, so the desired isomorphism follows immediately from
(\ref{E2gives}) and (\ref{H1W}).
The analogous result when $B={\bf P}^1$ and $G=E_6, E_7$ was proved by
Kanev [\ref{K1}]. The result in case $G=E_8$ and $B$ is 1-dimensional is 
announced in [\ref{FMW2}].

Next, the discrete part, i.e. the group of connected 
{\em components} of  $Prym_{\Lambda}(\widetilde{B})$. 
From (\ref{cl}) we have:
\beqa\label{compos}
0 \ra Comp(Prym_{\Lambda}(\widetilde{B})) \ra 
Comp(Hom_W(\Lambda, Pic(\widetilde{B}))) \ra H^2(W,T).
\eeqa
Now from the long-exact sequences of $W$-cohomology of the two short exact 
sequences: 
$$0 \ra Pic^0(\widetilde{B}) \ra 
Pic(\widetilde{B}) \ra H^{11}(\widetilde{B},{\bf Z}) \ra 0$$ 
and 
$$0 \ra H^1(\widetilde{B}), {\bf Z}) \ra 
H^1(\widetilde{B}), {\bf R}) \ra Pic^0(\widetilde{B}) \ra 0$$
we deduce:
\beqa\label{compareto}
0 \ra  H^1(W,Hom(\Lambda,H^1(\widetilde{B},{\bf Z})))
\ra  Comp(Hom_W(\Lambda, Pic(\widetilde{B}))) \ra \\
\ra Hom_W(\Lambda, H^{11}(\widetilde{B},{\bf Z})) 
\ra H^1(W,Hom(\Lambda,Pic^0(\widetilde{B}))). \nonumber
\eeqa
Noting the isomorphism of $H^1(W,Hom(\Lambda,Pic^0(\widetilde{B})))$ with 
$H^2(W,Hom(\Lambda, H^1(\widetilde{B},{\bf Z})))$, we can map the entire
sequence (\ref{compareto}) into (\ref{E1gives}). The first and last terms match 
exctly, while between the third terms we get an inclusion. We have therefore 
identified the discrete data
$ Comp(Prym_{\Lambda}(\widetilde{B})) = 
Comp(Hom_W(\Lambda, Pic(\widetilde{B})))$ 
with the subspace of  $H^2_W$ which is of Hodge type $(1,1)$. 
From (\ref{E2gives}), 
we see that  the group of components of the Deligne cohomology, 
$H^{22}_0(U,{\bf Z})$, is thus identified with a subgroup of finite index in 
$Comp(Prym_{\Lambda}(\widetilde{B}))$. We would get an actual isomorphism if we 
had an isomorphism in (\ref{E2gives}); for instance, this would follow if we knew that 
$H^2(W,Hom(\Lambda,{\bf Z}[W]))=0$. We have also made the assumption 
that  $H^2(W,T) = 0$. If this turns out to be non-zero, there could be a finite 
discrepancy, though this seems rather unlikely: heuristically, the group of 
components would  be reduced according to (\ref{compos}), while $H^2_W$ 
would be similarly reduced due to the non-zero $_1E^{30}$ term, and the 
effects are likely to cancel so that the isomorphism could be preserved. In fact, 
in case $G=E_6, E_7, E_8$ and $B= {\bf P}^1$, one indeed gets an isomorphism, 
according to [\ref{FMW2}].

To summarize, we have proved that the continuous part of the twisting data is
given by the relative intermediate Jacobian $J^3(U/B)$, and that the discrete 
part of the twisting data contains $H^{22}_0(U,{\bf Z})$ as a subgroup of finite 
index. This finite index is zero if, as we expect, the two group cohomologies 
$H^2(W,T)$ and $H^2(W,Hom(\Lambda,{\bf Z}[W]))$ both vanish.

\section{F-theory considerations}

{\bf 4.1 Transition to F-theory}

In the representation of the bundle via the del Pezzo construction respectively
in the stable degeneration on the $F$-theory side the data are already
in a form appropriate
to comparison. In the case of $E_8$ bundles one has
just to blow down the section of the $dP_9$ fibre on the $F$-theory side
to get the $dP_8$ fibre of the heterotic side showing the relation of
the cohomologies and hence of the intermediate jacobians.

Now for a bundle of group
$H\neq E_8$ the section $\theta :B\rightarrow X^4$ of 
$G$-singularities in the $F$-theory setup corresponds\footnote{We assume
$G$ to be simply-laced; otherwise one would have to consider the 
monodromy operation on the vanishing cycles of a corresponding
$F$-theory singularity, which are realized as outer automorphisms of 
the Dynkin diagram of the $F$-theory singularity leaving the 
non-simply-laced quotient as unbroken gauge group 
[\ref{AG}],[\ref{BIKMSV}].}
to having a bundle
of unbroken gauge group $G$, i.e. an $H$ bundle where $H$ is the 
commutant of $G$ in $E_8$, over the heterotic Calabi-Yau $Z$ respectively
having a section $s :B\ra {\cal W}_H={\cal M}_{Z/B}$ (at least locally
over the dense open subset of $B$ over which the fibres correspond to
semistable bundles) or, as the fibre of ${\cal W}_H$ over $b\in B$ 
parametrizes the corresponding del Pezzo surfaces, a bundle
$W_H^{het}\ra B$ of del Pezzo surfaces $dP_H$ fibered over $B$. We would 
like to see that the factor 
whose split off from $H^{2,1}(X^4)$ is caused by the 
local data along $\theta (B)$ is captured by $H^{2,1}(W_H^{het})$. The 
question is local in the $dP$ fibre and global along B. So 
locally at $\theta $, i.e. at the singularity along $B$, the picture in 
the $K3$ fibre of $X^4\ra B$ respectively the $dP$ fibre on the heterotic side
is the same if one considers heterotically actually a $dP_8$ 
fibration with $G$ singularity instead of the $dP_H$ fibration: 
this can be done as
we have an $ADE$ system of rational (-2) curves lying
in $H^{1,1}(K3,{\bf Z})$ as well as in $H^{1,1}(dP_8,{\bf Z})^{\bot_F}$
(in the case of the E-series, say; 
$F$ the elliptic curve representing the ample anticanonical divisor); 
note that the complex structures for $dP_H$ are given by 
homomorphisms $H^{1,1}(dP_H,{\bf Z})^{\bot_F} \ra F$
and the complex structures for $dP_8$ keeping the $G$ singularity
are similarly given by the corresponding homomorphisms for $dP_8$ mapping
the $G$ system of rational (-2) curves to zero (, i.e. they essentially
describe a mapping for the $H$-part). The matching of the cohomologies
gives the matching of the intermediate jacobians.

Note that as far as complex structure deformations are concerned
the distribution into deformations of $Z$ respectively those deformations
$H^1(W_i,T_{W_i}\otimes {\cal O}(-Z))\cong H^{3,1}(W_i)$ 
of $W_i$, which preserve $Z$, reflects the well known distribution of the
deforming monomials of the defining $F$-theory equation for $X^4$ 
between those which are "middle-polynomials" and the rest. \\ \\
{\bf 4.2 A remark on the four-flux}

We close with a remark on the 
four-flux.
We saw above how the identification of the necessary number of 
three-branes (from tadpole cancellation) is modified in presence of the 
four-flux [\ref{DM}] and how this is reflected on the heterotic side.
As a further example of the mutual interference of consistency conditions
we will see now how the fact that we take the space-time filling branes
into consideration refines the usual four-flux quantization condition.

Let us consider a $N=2$ compactification to 3D of $M$-theory on a 
Calabi-Yau fourfold $X$; this corresponds, in a certain limit for
elliptically fibered $X$, to a $N=1$ F-theory compactification to 4D.
In [\ref{W4fl}] it was shown that one has as quantization law for
the four-flux $G=\frac{1}{2\pi}dC$ not the naive integrality of 
$G$ but
$G=\frac{c_2}{2}+\alpha $
where $\alpha \in H^4(X,{\bf Z})$, i.e. in case $c_2$ is not even one
is not free in the decision to turn on four-flux or not 
or, to formulate it differently, the possible '0-value' has
changed from 0 to $\frac{c_2}{2}$. We will
see below that to achieve the wanted amount of supersymmetry in a
consistent compactification, $\alpha$ is even further restricted than 
to be merely an integral class. For example the easiest way to solve the 
congruence,
namely to put $G=\frac{c_2}{2}$, is thereby (besides by possible
non-primitiveness) ruled out if one wants to achieve $N=1$ supersymmetry,
i.e. there is actually no such thing in general as some simplest
(and may be even shifted) '0-value'.

 First recall that one has from the 
necessity 
of tadpole-cancellation, that a number $n_3$ of spacetime-filling branes 
has to be turned on [\ref{SVW}]
$\frac{\chi}{24}=n_3$ (here and in the following we have to assume
$\chi \geq 0$).
If one actually includes both degrees of freedom one finds [\ref{DM}]
\beqa
\frac{\chi}{24}=n_3+\frac{1}{2}\int G\wedge G
\eeqa
Furthermore we have [\ref{SVW}]
$\int c_2^2=480+\frac{\chi}{3}$
so the Euler number completely cancels out and one has 
\beqa
n_3=-60-\frac{1}{2}\int \alpha ^2+\alpha c_2
\eeqa
and, because of $n_3\ge 0$ to keep the supersymmetry [\ref{SVW}],
\beqa
\int \alpha ^2+\alpha c_2 \leq -120
\eeqa
On the other hand one has from the self-duality of $G$ that 
$G^2=\int G \wedge G\geq 0$ so that we get finally the following bounds
\beqa
-120 -\frac{\chi}{12}\leq \alpha ^2+\alpha c_2 \leq -120
\eeqa

For example one has for 
$\alpha \in c_2^{\bot}$ that $\alpha ^2 \leq -120$,
which as remarked especially rules out $G=\frac{c_2}{2}$; or,
to give an example where a condition $\alpha c_2 \neq 0$ occurs,
for $\alpha =D^2$ with a divisor which contributes to the superpotential
(i.e. $-\frac{1}{24}\alpha c_2=\chi (D, {\cal O}_D)=1$, cf. [\ref{W}] 
and [\ref{DGW}]) one finds $D^4\leq -96$, i.e. again a contradiction 
as the left
hand side vanishes for the vertical $D$ (more generally you find for 
$\alpha=D^2$ the condition $\chi (D,{\cal O}_D)\ge 5$ or equivalently 
$\chi ({\cal O}(D))\le -5$).
Note conversely that for a divisor $D$ contributing to the 
superpotential $D^2\cdot \alpha=12$ from
$G|_D=0$ [\ref{DMW}].\\ 

It is our pleasure to thank Paul Aspinwall, Pierre Deligne, Victor Ginzburg, 
Peter Mayr, Dave Morrison, Tony Pantev, Savdeep Sethi and Edward Witten 
for useful discussions.

\bigskip
\bigskip

\appendix

{\Large {\bf Appendix}}

\section{Root systems and del Pezzo surfaces}

Useful references are [\ref{MS}],[\ref{DKV}],[\ref{Ma}],[\ref{Dem}] in
general and especially [\ref{K1}],[\ref{K2}].

First one makes the transition (cf. below)
from the lattice $L=Q(R)\subset {\bf h}^*$
to its extension $N(L,\lambda)=L\oplus l{\bf Z}$ 
related to the embedding of the root system
into $H^{1,1}(dP,{\bf Z})$, where in addition to the blowing up
classes also the line class of the original $P^2$ lives 
(or equivalently,
as a certain linear combination with the other classes, 
the further class $f=-K$).

The fact that $C\rightarrow B$ is modelled 
($W$-equivariantly with fibre the Weyl orbit $W\lambda$)
on (2.8) translates to the
representation of the Prym-Tyurin variety as the image $P(C,i)$ of
the endomorphism $i-1$ of $J(C)$ coming from the correspondence
$G=D-\Delta$ ($\Delta$ the diagonal, so $i$ coming from $D$) 
given (outside the ramification locus) by
\beqa
G(x)=\sum_{j=1}^d(x,l_j)l_j
\eeqa
(using the identification of the fibre 
$W\lambda\cong Wl=\{(l_j)_{j=1,\dots d=|R|}\}$; for the 
$D$ correspondence the sum goes over $l_j\neq x$, it is of degree
$n=deg_CD(x)=(l,\sum_jl_j)+1=d+1$;
the occurring scalar products are for example $0,\pm 1$ in the E-series 
up to $E_6$ and include further $2$ and even $3$ for $E_7$ respectively. 
$E_8$).

Generalizing the involution $i^2=1$, i.e. $(i-1)(i+1)=0$,
of the D-series (cf. below), which leads to the
ordinary Prym $J^-(C)$, one has $(i-1)(i+q-1)=0$ with 
$q=-(\lambda | \lambda)\frac{d}{r}\in {\bf N}$; so $q=1\leftrightarrow
D=0\leftrightarrow W\lambda=W\omega_1$ in the A-series, $q=2$
for $D_r$ and $W\omega_1$ orbit and generally 
$q=2\frac{|R|}{r}=2h$ for $\lambda=\tilde{\alpha}$, the maximal root
($W\lambda=R$, $h$ the Coxeter number).

As pointed out above, one makes first the transition from the lattice 
$L=Q(R)\subset {\bf h}^*$
to its extension $N(L,\lambda)=L\oplus l{\bf Z}$ 
related to the embedding into the extended root system respectively to
the one into $H^{1,1}(dP,{\bf Z})$.

The W-operation is linearly extended by $w(kl)=k(w\lambda-\lambda)+kl$.
With $\kappa=\sum_ww(l)$ one has 
$N(L,\lambda)_{\bf Q}=L_{\bf Q}\oplus \kappa {\bf Q}$, 
$l=\lambda+c\kappa$
and the relation of orbits $W\lambda=(\lambda_i)_i$ and 
$Wl=(l_i=\lambda _i+c\kappa)_i$, so that the projection on the first
summand $p:N(L,\lambda)_{\bf Q}\rightarrow L_{\bf Q}$ gives a 
W-equivariant bijection $Wl\rightarrow W\lambda$. Furthermore the 
symmetric, bilinear, W-invariant, negative definite form 
$(\cdot |\cdot )$
on $L_{\bf Q}$ extends to a unique symmetric, bilinear, W-invariant
form $(\cdot ,\cdot )$ on $N(L,\lambda)_{\bf Q}$ with 
$(\alpha,l)=(\alpha,\lambda)$ for $\alpha \in L_{\bf Q}$ and $(l,l)=-1$.
One has $(l_{\alpha},l_{\beta})=(\alpha |\beta)+1$ and of course
$(\alpha |\beta)\in o,\pm 1$ apart from $(\alpha |\pm \alpha)=\mp 2$.
\\ \\
$\underline{{\bf A_r}}$ $\;$ $R=\{(\epsilon_i-\epsilon_j)\},i\neq j$ and
$i,j=1\dots r+1$ of basis $\alpha_k=\epsilon_k-\epsilon_{k+1}$ and
fundamental weights $\omega_i$; for $\lambda=\omega_1$ one has 
$W\lambda=\{(\lambda_i)\}, 
\lambda_i=\epsilon_i-\frac{1}{r+1}\sum_{i=1}^{r+1}\epsilon_i$, so $d=r+1,
W\cong S_d$ and $N(R,\lambda)=\oplus_{i=1}^d \epsilon_i{\bf Z}$ and the
elements $(\epsilon_i)$ of the orbit $Wl$ are 'disentangled' with 
respect to $(\cdot ,\cdot )$: $(\epsilon_i, \epsilon_j)=\delta_{ij}$, 
i.e. $D=0$ and $P(C,i)=J(C)$\\
$\underline{{\bf D_r}}$ $\;$ $R=\{(\pm\epsilon_i\pm\epsilon_j)\},i\neq j$
and $i,j=1\dots r$ of basis $\alpha_i=\epsilon_i-\epsilon_{i+1},i=1,\dots
r-1,\alpha_r=\epsilon_{r-1}+\epsilon_r$; for $\lambda=\omega_1$ one has
$W\lambda=\{(\pm\epsilon_i)\}_{i=1,\dots r}, d=2r$, and correspondingly
$Wl=\{(l_{\pm i})\}_{i=1,\dots r}$ (where 
$l_{\pm i}\leftrightarrow \epsilon_{\pm i}:=\pm\epsilon_i$); now
$(l_i,l_j)=0$ for $i\neq j$, $(l_i,l_{\pm i})=\mp 1$ and $Dx_i=x_{-i}$,
i.e. $i$ is an involution and $P(C,i)=J^-(C)$, the ordinary Prym\\
$\underline{{\bf E_r}}$ $\;$ take $\lambda=\omega_r$ 
(for $4\leq r \leq 7$
miniscule, $\omega_8=\tilde{\alpha}$ quasi-miniscule (there exist no
miniscule weights for $E_8$)); $d=|R|, q=2h$; the description will be 
continued below in the $H^{1,1}(dP,{\bf Z})$ language\\

Now we will come to the description of the root systems via the 
exceptional classes in $H^{1,1}(dP,{\bf Z})$. Let $L_i$ denote the 
exceptional classes from the blow up process in $dP_k$, $h$ the class 
of the line from $P^2$.\\ \\
$\underline{{\bf A_r}}$ $\;$ $\epsilon_i\rightarrow L_i$ extends to 
$N(R,\lambda)\oplus h{\bf Z}\cong H^{1,1}(dP_{r+1},{\bf Z})$ as an
orthogonal decomposition\\
$\underline{{\bf D_r}}$ $\;$ take the representation of $dP_{r+1}$
as Hirzebruch surface $F_n$ blown up in $r$ points lying on different
fibers, denote the two $P^1$ in each of the $r$ special fibers of type
$A_2$ by $L_{\pm i}$ ($i=1,\dots r$) of classes $l_{\pm i}$ and by $f$ 
the fibre class ($f=l_i+l_{-i}$); $f^{\bot}$ is the sublattice 
generated by the $l_i$ ($\pm i=1,\dots r$), $(f+K)^{\bot}$ is generated 
by the root system $R=\{(l_{\pm i}-l_{\pm j})\},\pm i,\pm j=1, \dots r,
i\neq \pm j$ of type $D_r$ and $f^{\bot}=N(R,\omega_1)$ (W-equivariantly)
making $\{(l_{\pm i})\}$,$i=1,\dots r$ correspond to $Wl$\\
$\underline{{\bf E_r}}$ $\;$ one has with $f=-K$ that 
$\{\alpha\in f^{\bot}|\alpha^2=-2\}\cong E_r$ 
of basis $(L_{i+1}-L_i)$,$i=1,\dots r-1$, representing the line in the
$E_r$ Dynkin diagram and $L_1+L_2+L_3-L_0$ the special node not lying
on the line. Furthermore
$N(R,\omega_r)\cong H^{1,1}(dP_r,{\bf Z})$ (W-equivariantly) being 
$id$ on $Q(R)$ and sending $l$ to $L_r$ 
and $Wl$ to the set $I_r$ of (-1)-curves $c$ with $c^2=-1,c\cdot f=1$
(and $\kappa=\sum_{i=1}^rl_i$ to $cK=\sum_{j=1}^rL_j$) (one has 
$E_5=D_5,E_4=A_4,E_3=A_1\times A_2$).\\ \\

\section*{References}
\begin{enumerate}

\item
\label{FMW}
R. Friedman, J. Morgan and E. Witten, {\it Vector Bundles and 
F-Theory}, Commun. Math. Phys. {\bf 187} (1997) 679, hep-th/9701162.

\item
\label{D}
R.Y. Donagi, {\it Principal bundles on elliptic fibrations}, 
Asian J. Math. {\bf1} (1997), 214-223, alg-geom/9702002.

\item
\label{D2}
R.Y. Donagi, {\it  Spectral covers, in: Current topics in complex algebraic 
geometry}, MSRI pub. {\bf 28} (1992), 65-86, alg-geom 9505009.

\item
\label{BJPS}
M. Bershadsky, A. Johansen, T. Pantev and V. Sadov, {\it On 
four-dimensional Compactifications of F-theory}, hep-th/9701165.

\item
\label{FMW2}
R. Friedman, J. Morgan and E. Witten, {\it Principal G-Bundles over
elliptic curves}, alg-geom/9707004.

\item
\label{V}
C. Vafa, {\it Evidence for $F$-theory}, Nucl. Phys. {\bf B 469} (1996) 
403, hep-th/9602022.

\item
\label{SVW}
S. Sethi, C. Vafa and E. Witten, {\it Constraints on Low-Dimensional 
String Compactifications}, hep-th/9606122.

\item
\label{MV}
D. Morrison and C. Vafa, {\it Compactifications of F-theory on Calabi-Yau
Threefolds I, II}, Nucl. Phys. {\bf B 473} (1996) 74; ibid. {\bf B 476}
(1996) 437.

\item
\label{C}
G. Curio, {\it Chiral Multiplets in $N=1$ dual string pairs}, 
hep-th/9705197.

\item
\label{AC}
B. Andreas and G. Curio, {\it Three-branes and five-branes in $N=1$
dual string pairs}, hep-th/9706093.

\item
\label{AM}
P.S. Aspinwall and D.R. Morrison, {\it Point-like Instantons on
K3 Orbifolds}, hep-th/9705104.

\item
\label{W4fl}
E. Witten, {\it On Flux Quantization in M Theory and the Effective 
Action},
hep-th/9609122.

\item
\label{BB}
K. Becker and M. Becker, {\it {\cal M}-theory on eight-manifolds}, 
hep-th/9605053.

\item
\label{FMW3}
R. Friedman, J. Morgan and E. Witten, {\it Vector Bundles over 
elliptic fibrations}, alg-geom/9709029.

\item
\label{Gr}
A. Grassi {\it Divisors on elliptic Calabi-Yau 4-folds and the 
superpotential in $F$-theory}, alg-feom/9704008.

\item
\label{KLRY}
A. Klemm, B. Lian, S.-S. Roan and S.-T. Yau, {\it Calabi-Yau fourfolds
for $M$- and $F$-theory compactifications}, hep-th/9701023.

\item
\label{ACL}
B. Andreas, G. Curio and D. L\"ust, {\it $N=1$ dual string pairs and 
their massless spectra}, hep-th/9705174.

\item
\label{DM}
K. Dasgupta and S. Mukhi, {\it A Note on Low-Dimensional String
Compactifications}, Phys.Lett. {\bf B 398} (1997) 285, hep-th/9612188. 

\item
\label{KMV}
S. Katz, P. Mayr and C. Vafa, {\it Mirror symmetry and exact solution
of 4D $N=2$ Gauge theories I}, hep-th/9706110.

\item
\label{G}
O. Ganor, {\it Toroidal compactification of heterotic 6D
non-critical strings down to four dimensions}, hep-th/9608109.

\item
\label{CCML}
G. Lopes Cardoso, G. Curio, T. Mohaupt and D. L\"ust, {\it On the 
duality between the heterotic string and F-theory in 8 dimensions}, 
Phys. Lett. {\bf B389} (1996) 479, hep-th/9609111.

\item
\label{HM}
J. Harvey and G. Moore, {\it Exact gravitational threshold correction
in the FHSV model}, hep-th/9611176.

\item
\label{L}
W. Lerche, {\it BPS states, weight spaces and vanishing cycles}, 
Proceedings of "Strings 95", Nucl. Phys. Proc. Suppl. {\bf 46} (1996) 122,
hep-th/9507011.

\item
\label{KLMVW}
A. Klemm, W. Lerche, P. Mayr, C. Vafa and N. Warner, {\it Self-Dual 
Strings and N=2 Supersymmetric Field Theory}
Nucl. Phys. {\bf B477} (1996) 746, hep-th/9604034.

\item
\label{MS}
D.R. Morrison and N. Seiberg, {\it Extremal transitions and 
five-dimensional supersymmetric field theories}, hep-th/9609070.

\item
\label{DKV}
M.R. Douglas, S. Katz and C. Vafa, {\it Small instantons, del Pezzo 
surfaces and type I' theory}, hep-th/9609071.

\item
\label{Ma}
Yu.I. Manin, {\it Cubic Forms: Algebra, Geometry, Arithmetic}, 
North-Holland, Amsterdam, New York, 1986.

\item
\label{Dem}
M. Demazure, {\it Surfaces de del Pezzo}, Seminaire sur les Singularities
des Surfaces, Lecture Notes in Mathematics vol 777, Springer-Verlag, 1980.

\item
\label{DD}
R.Y. Donagi, {\it Decomposition of Spectral covers}, 
Journees de Geometrie Algebrique d'Orsay, Asterisque 218
(1993) 145.

\item
\label{K2}
V. Kanev, in "Abelian Varieties" (W. Barth, K. Hulek and H. Lange, eds.)
Proceedings of the International Conference (Egloffstein 1993), de
Gruyter, Berlin- New York, 1995.

\item
\label{K1}
V. Kanev, Annali di Matematica pura ed applicata (IV), Vol. CLIV 
(1989) 13.

\item
\label{AG}
P. Aspinwall and M. Gross, {\it The $SO(32)$ Heterotic String on a K3
surface}, hep-th/9605131.

\item
\label{BIKMSV}
M. Bershadsky, K. Intrilliator, S. Kachru, D.R. Morrison, V. Sadov and
C. Vafa, {\it Geometric Singularities and Enhanced Gauge Symmetries},
hep-th/9605200.

\item
\label{W}
E. Witten, {\em Nonperturbative Superpotentials in String Theory},
Nucl. Phys. {\bf B 474} (1996) 343, hep-th/9604030.

\item
\label{DGW}
R. Donagi, A. Grassi and E. Witten, {\it A nonperturbative superpotential
with $E_8$ symmetry}, hep-th/9607091.

\item
\label{DMW}
M.J. Duff, R. Minasian and E. Witten, {\it Evidence for 
Heterotic/Heterotic Duality}, Nucl. Phys. {\bf B465} (1996) 413, 
hep-th/9601036.

\end{enumerate}
\end{document}